\documentclass[sigplan,screen]{acmart}
\startPage{1}

\setcopyright{acmlicensed}
\acmPrice{15.00}
\acmDOI{10.1145/3519939.3523705}
\acmYear{2022}
\copyrightyear{2022}
\acmSubmissionID{pldi22main-p334-p}
\acmISBN{978-1-4503-9265-5/22/06}
\acmConference[PLDI '22]{Proceedings of the 43rd ACM SIGPLAN International Conference on Programming Language Design and Implementation}{June 13--17, 2022}{San Diego, CA, USA}
\acmBooktitle{Proceedings of the 43rd ACM SIGPLAN International Conference on Programming Language Design and Implementation (PLDI '22), June 13--17, 2022, San Diego, CA, USA}



\usepackage{booktabs} 
\usepackage{subcaption}

\usepackage[utf8]{inputenc}
\usepackage{algorithm}
\usepackage[noend]{algpseudocode}
\usepackage{paralist}
\usepackage{xspace}
\usepackage{algpseudocode}
\usepackage{tikz}
\usepackage{adjustbox}
\usepackage{amsthm}
\usepackage{multirow}
\usepackage{listings}
\usepackage{algorithmicx}
\algdef{SE}[DOWHILE]{Do}{doWhile}{\algorithmicdo}[1]{\algorithmicwhile\ #1}%
\usepackage{alltt}
\usepackage{fancyvrb}
\usepackage{enumitem}
\usepackage{mdframed}

\usepackage[most]{tcolorbox}

\usetikzlibrary{fit,decorations.pathreplacing,calligraphy,calc}

\newcommand{\ignore}[1]{}

\newcommand{\technique}{\lrsyn\xspace}

\newcommand{\leaveout}[1]{{}}

\newcommand{\hdataset}{\ensuremath{\mathcal{D}}}
\newcommand{\dataset}{\ensuremath{D}}
\newcommand{\doc}{\ensuremath{\mathsf{doc}}}

\newcommand{\cluster}{\ensuremath{C}}
\newcommand{\field}{\ensuremath{\mathsf{F}}}
\newcommand{\type}{\ensuremath{\mathsf{T}}}
\newcommand{\locations}{\ensuremath{\mathsf{Locs}}}
\newcommand{\location}{\ensuremath{\ell}}
\newcommand{\data}{\ensuremath{\mathsf{Data}}}
\newcommand{\annotation}{\ensuremath{\mathcal{A}}}
\newcommand{\aggregate}{\ensuremath{\mathsf{Agg}}}
\newcommand{\train}{\ensuremath{\mathsf{tr}}}
\newcommand{\extract}{\ensuremath{\mathsf{Extract}}}
\newcommand{\landmark}{\ensuremath{\mathsf{m}}}
\newcommand{\landmarks}{\ensuremath{\mathsf{M}}}
\newcommand{\locate}{\ensuremath{\mathsf{Locate}}}
\newcommand{\region}{\ensuremath{\mathsf{R}}}
\newcommand{\locationset}{\ensuremath{L}}
\newcommand{\blueprint}{\ensuremath{\mathsf{BP}}}
\newcommand{\bpvalue}{\ensuremath{\mathsf{b}}}
\newcommand{\dtr}{\ensuremath{\dataset_\train}}
\newcommand{\RProg}{\ensuremath{\mathsf{RProg}}}
\newcommand{\EProg}{\ensuremath{\mathsf{EProg}}}
\newcommand{\FullProg}{\ensuremath{\mathsf{Prog}}}
\newcommand{\fvalue}{\ensuremath{v}}

\newcommand{\ndsyn}{\ensuremath{\mathsf{NDSyn}}\xspace}
\newcommand{\lrsyn}{\ensuremath{\mathsf{LRSyn}}\xspace}
\newcommand{\resyn}{\ensuremath{\mathsf{LRSyn}}\xspace}
\newcommand{\afr}{\ensuremath{\mathsf{AFR}}}
\newcommand{\fx}{\ensuremath{\mathsf{ForgivingXPaths}}}

\newcommand{\Examples}{\mathsf{Ex}}

\newcommand{\lleq}{\ensuremath{\preccurlyeq}}
\newcommand{\RegionSpec}{\ensuremath{\mathsf{RegionSpec}}}
\newcommand{\ExtractionSpec}{\ensuremath{\mathsf{ValueSpec}}}
\newcommand{\LandmarkCandidates}{\ensuremath{\mathsf{LandmarkCandidates}}}
\newcommand{\parentHops}{\ensuremath{\mathsf{parentHops}}}
\newcommand{\siblingHops}{\ensuremath{\mathsf{siblingHops}}}
\newcommand{\BoxSummary}{\ensuremath{\mathsf{BoxSummary}}}
\newcommand{\textbox}{\ensuremath{\mathsf{box}}}

\newcommand{\encr}{\ensuremath{\mathsf{EncRgn}}}

\newcommand{\fpd}{\ensuremath{\delta}}

\newcommand{\JointClusterAndLandmark}{\text{\textsc{InferLandmarksAndCluster}}}
\newcommand{\SynthesizeExtractionProgram}{\text{\textsc{SynthesizeExtractionProgram}}}

\newcommand{\LL}{\mathcal{L}}

\theoremstyle{definition}

\newcommand{\highlightbox}[1]{
\begin{mdframed}[backgroundcolor=green!10,skipabove=1ex,skipbelow=1ex]
#1
\end{mdframed}
\vspace{1ex}
}

\newcommand{\lrsynprogram}[6]{
\begin{alltt}
\textbf{Landmark}: #1 \\
\textbf{Region program:}~$\mathsf{parentHops}: #2,~\mathsf{siblingHops}: #3$ \\
\textbf{Blueprint}: #4\\
\textbf{Value program}: \\
\hspace{8ex}\textbf{CSS selector}: $#5$ \\
\hspace{8ex}\textbf{Text program}: #6
\end{alltt}
}

\AtEndPreamble{%
\theoremstyle{acmdefinition}
\newtheorem{remark}[theorem]{Remark}}

\renewcommand{\paragraph}[1]{\smallskip\noindent\textbf{#1.}}
\renewcommand{\subparagraph}[1]{\smallskip\noindent\emph{#1.}}

\begin{document}

\title{Landmarks and Regions: A Robust Approach to Data Extraction}

\author{Suresh Parthasarathy}
\email{supartha@microsoft.com}
\affiliation{\institution{Microsoft}\city{London}\country{UK}}

\author{Lincy Pattanaik}
\email{lincy.pattanaik@microsoft.com}
\affiliation{\institution{Microsoft Research}\city{Bangalore}\country{India}}

\author{Anirudh Khatry}
\email{t-ankhatry@microsoft.com}
\affiliation{\institution{Microsoft Research}\city{Bangalore}\country{India}}

\author{Arun Iyer}
\email{ariy@microsoft.com}
\affiliation{\institution{Microsoft Research}\city{Bangalore}\country{India}}

\author{Arjun Radhakrishna}
\email{arradha@microsoft.com}
\affiliation{\institution{Microsoft}\city{Redmond}\country{United States}}

\author{Sriram K. Rajamani}
\email{sriram@microsoft.com}
\affiliation{\institution{Microsoft Research}\city{Bangalore}\country{India}}

\author{Mohammad Raza}
\email{moraza@microsoft.com}
\affiliation{\institution{Microsoft}\city{Redmond}\country{United States}}

\renewcommand{\shortauthors}{S. Parthasarathy et al.}

\begin{abstract}
We propose a new approach to extracting data items or field values from semi-structured documents. Examples of such problems include extracting passenger name, departure time and departure airport from a travel itinerary, or extracting  price of an item from a purchase receipt.
Traditional approaches to data extraction use machine learning or program synthesis to process the whole document to extract the desired fields. Such approaches are not robust to format changes in the document, and the extraction process typically fails even if changes are made to parts of the document that are unrelated to the desired fields of interest. We propose a new approach to data extraction based on the concepts of {\em landmarks} and {\em regions}. Humans routinely use landmarks in manual processing of documents to zoom in and focus their attention on small regions of interest in the document. Inspired by this human intuition, we use the notion of landmarks in program synthesis to automatically synthesize extraction programs that first extract a small region of interest, and then automatically extract the desired value from the region in a subsequent step.  We have implemented our landmark based extraction approach in a tool \resyn, and show extensive evaluation on documents in HTML as well as scanned images of invoices and receipts. Our results show that the our approach is robust to  various types of format changes that routinely happen in real-world settings.
\end{abstract}

\maketitle

\section{Introduction}
Extracting data from semi-structured documents is an important and pervasive problem, which arises in many real-world situations. 
Our goal is to automatically synthesize extraction programs that can extract relevant fields  of interest from \emph{formed semi-structured documents}.
We use the term formed document to represent a broad category of documents,
ranging from invoices, receipts, business cards, booking emails, etc.
While loosely defined, formed document collections have some key common
characteristics:
\begin{itemize}
\item
  They are often \emph{machine-to-human}, i.e., automatically generated by
    software, but are meant for consumption by humans. Usually, many documents are generated with the same format, which makes them amenable to programmatic data extraction. 
\item
  They are \emph{heterogeneous}, i.e., each kind of document can be in
    many ad-hoc formats.
    For example, while extracting information from flight reservation emails, each airline follows a different format, and the same airline may change formats over time. 
\item 
They are \emph{continuously evolving} with new document formats being
    added often. 
    Hence, a system for extracting documents is not a one-time task ---the system needs to be robust to evolution, and needs to be 
    routinely updated and maintained.
\end{itemize}
A \emph{robust} extraction system should handle the heterogeneous data formats at scale, with minimal human intervention and be efficient at run-time.

\leaveout{

Extraction is a well-studied problem in both the machine learning (ML) and programming language (PL) communities. Dominant approaches in the ML community include training sequential or transformer-based models ~\cite{}. Approaches in the PL community include synthesizing extraction programs based on examples~\cite{gulwani-et-al}, with several ideas to make use of structure and data signals~\cite{Predictive-synthesis,etc}. ML approaches require large amounts of labelled training data and lack interpretability of the learnt models, while PL approaches tend to learn specialized programs that may overfit the training examples and are unable to capture the vast amount of noise or possible variations in formats. Typically, even minor changes to the format of the input documents results in extraction failures.
Recently, there have been works to combine the ML and PL paradigms and come up with interesting extraction approaches which take the best of both these worlds~\cite{HDEF, DilligPLDI21}.
The combined approaches use program synthesis to provide interpretability and predictability, and rely on ML to provide robustness and adaptability when the formats change. 
Despite previous work, program synthesis approaches to extraction suffer from lack of robustness.
}

\begin{figure*}[th]
\centering
\begin{minipage}{5.0cm}
\frame{\includegraphics[width=5.0cm]{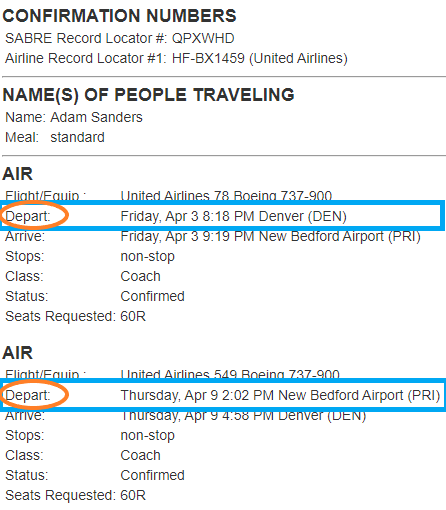}}
\vspace{-3ex}
\captionof*{figure}{a. Flight booking email}
\end{minipage}
\quad
\begin{minipage}{5.1cm}
\frame{\includegraphics[width=5.1cm]{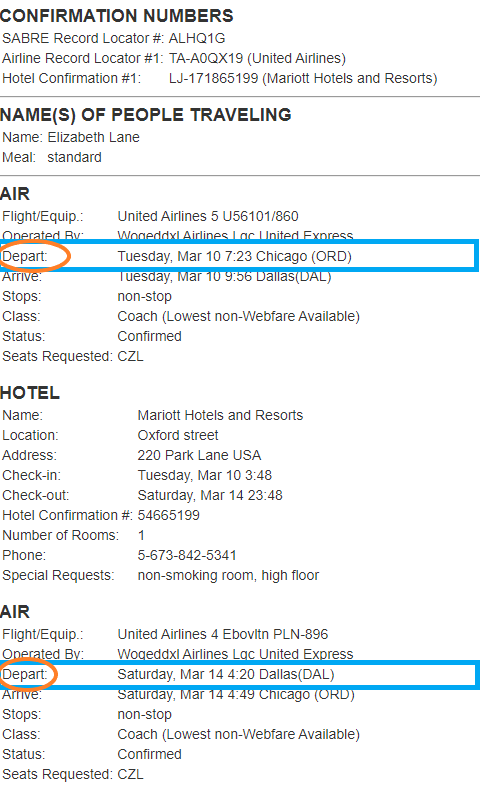}}
\vspace{-3ex}
\captionof*{figure}{b. Flight booking email}
\end{minipage}
\quad
\begin{minipage}{6.5cm}
\includegraphics[width=6.5cm]{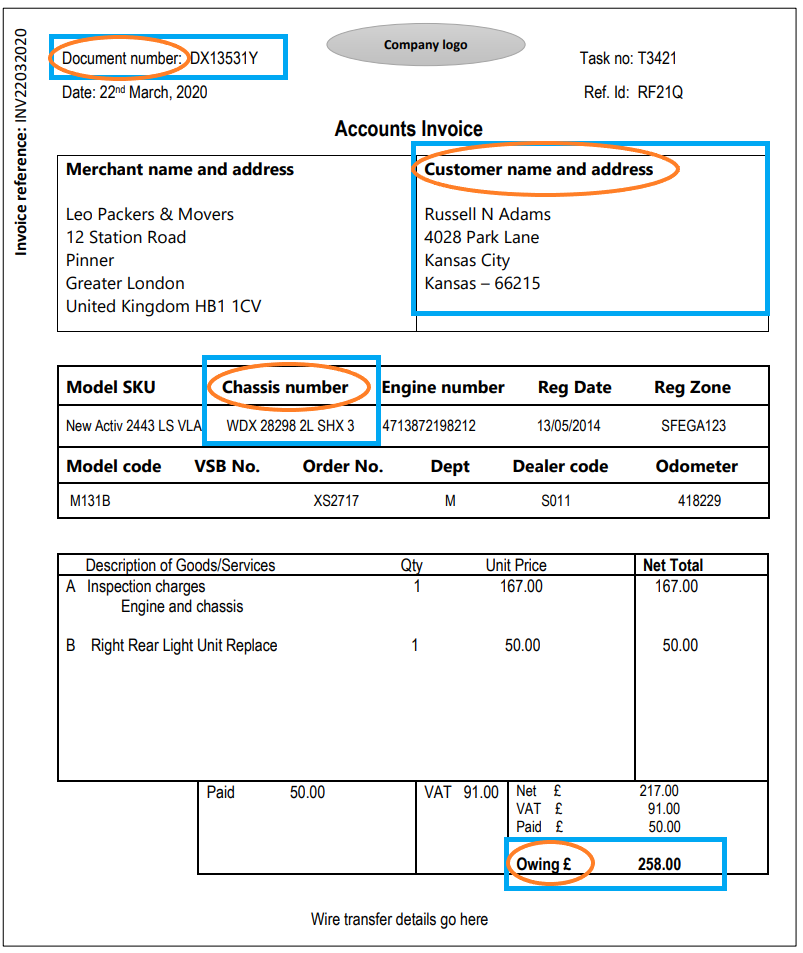}
\vspace{-3ex}
\captionof*{figure}{c. Accounts Invoice}
\end{minipage}
\vspace{-2ex}
\caption{Formed documents: Figure a. and b. are examples of flight reservation emails. Figure c. is an image of accounts invoice. Orange ellipses indicate landmarks and blue rectangles indicate ROIs}
\label{fig:formed_documents}
\vspace{-3ex}
\end{figure*}

Given this problem definition, our approach is inspired by how humans analyze
such complex and lengthy documents when looking for specific information: we
first narrow down the region that is relevant for the task at hand, and focus
only on that small region to extract and understand the desired information.
This inspires a key idea that we use in our approach, which is the notion of a
{\em landmark}.
In the literature, landmarks have been used to identify locations in document
which are ``nearby'' the value we desire to
extract~\cite{muslea1999hierarchical,yandrapally2014robust}.
%
As an analogy, if we want to locate a restaurant in a map, and we know that it
is near the train station, we can first locate the train station in the map, and
then the restaurant by identifying its location relative to the train station.
In this case, the train station is the landmark used to locate the restaurant.
In the case of documents, humans tend to use keyword phrases that occur in all
(or most) documents   near the locations of the desired field values as
landmarks.
For example, the phrase "Depart:" is a potential landmark to locate the value of
departure time in the travel emails shown in
Figure~\ref{fig:formed_documents}(a) and (b), and the phrase "Owing" is a
potential landmark to locate the value of total invoice amount in the invoice
shown in the right side of Figure~\ref{fig:formed_documents}(c).

\leaveout{
Landmarks are a form of data invariance present in all documents of a format. The idea of exploiting such data invariance  or specific regions for extraction is not novel in itself: algorithms proposed in~\cite{brin1998extracting, agichtein2000snowball}~use commonly reoccurring phrasal patterns for web extraction given a seed set; in the vision community, algorithms like R-CNN~\cite{girshick2014rich}~narrow down the region before extracting the objects from an image. The main contribution of this paper is to use landmarks in program synthesis to synthesize extraction programs that are robust to format changes in the input document. 
}

Landmarks are a form of data invariance present in all documents of a format.
The key idea in this paper is to use landmarks to decompose the field extraction
problem into two different sub-problems:
\begin{enumerate}
\item
Region extraction: This step takes the document as input, and produces a small region of interest as output,  guided by the landmark, such that the region of interest contains the landmark as well as the field values of interest.
\item
Value extraction: This step takes the small region of interest produced by the first step as input, and extracts the desired field values from the region.
\end{enumerate}
Since real-world scenarios have heterogeneous formats, we design our value
extraction to be conditional on both the landmark and the layout of  the
identified region of interest.
We formalize these notions and in a novel generic design of domain specific
languages which we call \emph{landmark-based DSLs}, which are designed for data
extraction using the landmark-based approach.
Such DSLs contain separate language fragments for the region and value
extraction steps which can be instantiated arbitrarily for different domains,
and we describe such instantiations and corresponding synthesis algorithms for
the different concrete domains of HTML and scanned image documents.

While the strategy of using landmarks and region layouts nicely structures the extraction task into well-defined sub-problems, the remaining question is how we can infer the landmarks and region layouts themselves. Landmarks capture redundancy across documents with similar formats, and hence landmarks can be detected by analyzing commonalities across documents with the same format. The formats themselves can be identified using clustering techniques that group similar documents together, and the same techniques can be applied at the region level to infer different region layouts. In this work we show how the two problems of landmark detection and clustering can be solved by a general algorithm that jointly infers clusters and landmarks in a hierarchical fashion. 

We have implemented this approach in a tool called $\resyn$, which is short for {\bf Landmark-based Robust Synthesis}. We present empirical results comparing the performance of \resyn\  with current approaches on datasets containing documents from HTML and scanned image domains.

To summarize, we make the following contributions: 
\begin{itemize}
\item
We use the concept of \textbf{landmarks and regions} for extracting
attributes from heterogeneous data formats and introduce the formal generic
class of landmark-based DSLs in which we can express robust extraction programs
that continue to work when formats change.
\item
We propose a joint clustering and landmark detection algorithm which  automatically clusters and infers landmarks on the given input data.
\item
We present concrete instantiations of landmark-based DSLs and corresponding synthesis algorithms in the particular domains of HTML and scanned document images.
\item
Our implementation, $\resyn$, is comparable to state-of-the-art when test data is from the same time period as training data, and significantly outperforms current approaches when test data is from a different time period than training data. In the HTML domain, \resyn\ is able to achieve near perfect F1 score of 1.0 in most cases. In the images domain, \resyn\ outperforms a released product in scenarios where test data has different formats when compared to training data. This gives evidence that $\resyn$ is robust to format changes that occur over time  in real-world scenarios.
\end{itemize}

\section{Overview}
\label{sec:overview}

Many existing data extraction techniques~\cite{lg2014,ij2019,AFR}, both from
research literature and those used industrially, are driven by \emph{global
document structure}.
For example, consider the confirmation email for a flight reservation in
Figure~\ref{fig:formed_documents}(a), and the task of extracting the flight departure
time from it.
The \ndsyn\ algorithm from the HDEF system~\cite{ij2019}~synthesizes the extraction program shown in Figure~\ref{lst:ndsynprogram}~for this task.
\
This program starts with extracting the "TBODY" elements in the HTML document; which are the only TBODY child in a table. This results in the various blocks, "AIR", "HOTEL", etc. Within each block, the program extracts the $6^{th}$ child from the end and within that node, it extracts the $2^{nd}$ child. If an extra block gets added to the document between the two "AIR" blocks, like the "HOTEL" block Figure~\ref{fig:formed_documents}(b), the program extracts the "Check-in" time from the "HOTEL" block as well in addition to the departure times. This program also fails for cases where the "AIR" blocks have more elements.

\begin{figure}
\small
\begin{alltt}
\textbf{CSS selector}: :nth-child(11) > TABLE >
                     TBODY:nth-child(1):nth-last-child(1) 
                     > :nth-last-child(6) > :nth-child(2)
\textbf{Text program}: Extract TIME sub-string
\end{alltt}
\caption{\ndsyn\ extraction program for Departure time}
\label{lst:ndsynprogram}
\end{figure}



In contrast, local structure based data extraction techniques focus more on the
structure of the document that is close to the value to be extracted~\cite{muslea1999hierarchical,yandrapally2014robust}.
Inspired by these approaches, we propose landmark-based synthesis which more
closely relates to the way humans scan through documents, and functions in a
local and compositional manner.
Let us walk through how a human might find the departure time in
document.
First, the human begins by scanning the document for keyword phrases---in this case,
say, the word \emph{AIR}.
Now, the human finds the first title \emph{AIR} and then, scans the section
corresponding to the keyword.
Recursively, they may then search for the keyword \emph{Depart:} and find the
corresponding text that contains the actual flight departure time.
Thus, humans typically navigate documents using keyword phrases that occur in all documents in the neighborhood of the field values desired.

Our landmark based synthesis algorithm \resyn\ mirrors the above workflow. We introduce the notions of
\emph{landmarks},  \emph{regions of interest} and \emph{blueprints} to formalize this workflow.
Landmarks correspond to keyword phrases used in the human workflow, regions of interest (\emph{ROIs}, for short)
correspond to the local and relevant document sections around the landmarks, and we use
blueprints to compare similarity of regions.
In the above example, the keywords \emph{AIR} and \emph{Depart} act as
landmarks and their corresponding sections act as regions of interest.

\resyn\ consists of two main components:
\begin{inparaenum}[(a)]
  \item Identifying document landmarks from the data, which also involves clustering documents which have roughly similar local structure in the neighborhood of the desired field values together
  \item Synthesizing extraction programs, which first zoom in on the region of interest given a document using the landmark as the anchor, and then extract the desired field value from the region of interest.
\end{inparaenum}
We illustrate each of these steps below using our machine-to-human (M2H) email dataset, which consists of ~$3500$ HTML flight
reservation emails from $6$ airlines.

\subsection{Jointly inferring landmarks and clusters}
\label{subsec:joint-infer-cluster}

\paragraph{Initial Clustering}
The first step in our technique is to separate the data-set into clusters that
correspond to different formats of emails.
For this, we use the notion of {\em blueprints}. 
For the HTML domain, the blueprint of a document (or a document fragment)
contains the tag structure and values that are common across documents. 
We compute an initial clustering by considering of whole documents, and using the closeness of blueprints as the distance metric between documents. The initial clustering produces a large number of very fine-grained clusters.
On the M2H data-set, our initial clustering produces ~$20$ \emph{head
clusters} along with many small \emph{tail clusters} totaling ~$70$
clusters altogether. 

\paragraph{Identifying Landmark Candidates}
For each cluster, we identify a landmark, which is an n-gram that appears in all documents in the cluster.
For each such landmark candidate (i.e., shared n-grams), we assign a score based
on two metrics:
\begin{inparaenum}[(a)]
  \item the distance between the landmark candidate and the field value to be
    extracted, and
  \item the number of nodes in the document region that encloses both the
    landmark candidate and the field value.
\end{inparaenum}
These two metrics capture the intuition that landmarks should be close to the
values being extracted.
For the email in Figure~\ref{fig:formed_documents}(a), the  top-scoring landmark candidates for extracting \emph{Departure Time} are "Depart" and "Arrive".

\paragraph{Regions of Interest and Re-clustering}
A contiguous region in the document that encloses the landmark candidate and the field value is called a \emph{region of interest}. We are interested in small regions of interest.
In Figure~\ref{fig:formed_documents} the landmark candidates are shown in orange ellipses, and the corresponding regions of interest are shown using blue rectangles.
We now re-define a distance metric between two documents as the minimum distance (over all
landmark candidates) between the blueprints of the corresponding regions of interest.
Using this revised distance metric, we iteratively merge clusters if the average distance between documents in the clusters is less than a threshold, till no further merging is possible.
As a result of such merging, formats with an advertisement section added, or existing sections rearranged would all be included in the same cluster, as the blueprint of the region of interest is invariant to such  changes that occur outside the region of interest. In the M2H data-set, iterative merging based on closeness of regions of interest results in less than $5$ clusters at a field level.

\subsection{Synthesizing Extraction Programs.}
With the new coarse-grained clusters, we synthesize $2$ sub-programs that
together with the blueprint of the region of interest forms the complete
extraction program.
The first sub-program, called {\em region extraction program}, takes as input the whole document $d_i$ and a landmark location $\ell$ as inputs, and produces a region of interest $R$ as the output. 
  In the case of HTML documents, the synthesized region extraction program starts with the landmark location and traverses the tree-structure of the HTML document and grows a region starting at the location of the landmark, and ensures that all locations of field values are included.
  As an example,for the airline itinerary example shown in Figure~\ref{fig:formed_documents}(a), in order to extract the value of the departure time, with "Depart:" as the landmark, the synthesized region extraction program is given by \emph{0 parent hops, 1 sibling hop}. The single sibling hop here implies that across all the training data, the extraction value lies $1$ sibling away from the landmark within the same parent node in the DOM. The \emph{parent hops} and \emph{sibling hops} are computed based on all the annotated training documents in the given cluster, and hence produce a large enough ROI  that includes the location of all the field values for extraction. We also compute the blueprint for this region, and use this blueprint during execution of the region extraction program (see below).
  The second sub-program, called {\em value extraction program}, takes the region produced by the region extraction program as input, and produces the desired field value as output. For HTML documents, we use the DSL and the corresponding synthesis algorithm from~\cite{raza2020web}. For the example in Figure~\ref{fig:formed_documents}(a), the synthesized region and value extraction programs, and the blueprint of the region of interest are shown in Figure~\ref{lst:lrsynprogram}.

\begin{figure}
\small
\lrsynprogram{Depart}{0}{1}{/TD}{nth-child(2)}{Extract TIME sub-string}
\caption{\lrsyn\ extraction program for Depart time}
\label{lst:lrsynprogram}
\end{figure}




%
During execution of the synthesized extraction program, we compute the blueprint of the ROI  calculated by the region extraction program and compare it with the blueprint of the ROI generated during synthesis. If the distance between these two blueprints is below a threshold value $t$, the regions from synthesis and inference times are "roughly similar", and we use this extracted program. Otherwise, we look for other extraction programs synthesized from other clusters, which better match the current document.

\section{Formed Document Extraction}
\label{sec:problem}
\label{sec:formalism}

In this section, we present the formed document problem definition, formalize the notions of landmarks and regions, and introduce the novel class of landmark-based DSLs. 

\subsection{Preliminaries and Problem Statement}

\paragraph{Documents and Locations}
We use the term \emph{document} to represent a single record of a dataset from which we are interested in extracting data.
We use the symbol $\doc$ to represent documents.
A document $\doc$ has a set of \emph{locations} $\locations(\doc)$ which can be
used to index into the document and look up values.
For a location $\location \in \locations(\doc)$, the function $\data[\location]$
returns the value of the data present in the location $\location$ of $\doc$.
\begin{example}
    \label{ex:locs}
    When dealing with HTML documents, the location are XPaths that retrieve
    elements in the HTML DOM document tree structure.
    The data value of a location is the concatenation of all the text elements
    in the DOM element.
\end{example}

\paragraph{Data-sets and Fields}
We model a heterogeneous dataset $\hdataset$ as a  tuple $(\dataset, \{
\cluster_0, \ldots, \cluster_n \})$ where:
\begin{inparaenum}[(a)]
\item $\dataset$ is a finite set of \emph{input documents} (or just \emph
    {documents} for short) and
\item $\{ \cluster_0, \ldots, \cluster_n \}$ is a partition of $\dataset$ into
    \emph{clusters}, i.e., $\bigcup \cluster_i = D$ and $\forall i \neq j .\;
    \cluster_i \cap \cluster_j = \emptyset$.
\end{inparaenum}
Each partition represents a \textit{similar} set of documents in terms of format.
The extraction framework has access to the inputs $\dataset$, but not the
partitioning.
Henceforth, we write ``dataset $\dataset$'' instead of ``heterogeneous
dataset $\hdataset$'' to denote that the exact partition of $\dataset$ into
clusters is not provided to us as input.

For a given dataset $\dataset$, a  \emph{field} $\field$  of type $\type$
is a partial function  $\field : D \not\to \type$ that maps documents to
values of type $\type$.
%
We implicitly assume that a field is either defined for all documents in a
cluster or is undefined for all documents in the cluster, i.e.,
$\forall \cluster_i . \forall \doc, \doc' \in \cluster_i. 
    \field(\doc) = \bot  \Leftrightarrow \field(\doc') = \bot$.
%
We say that $\field(\doc)$ is the {\em value of the field} $\field$
in $\doc$.
The type of a field can either be a primitive type such as integer or string or a
composite type such as a list of strings or set of integers.
Though we are interested in extracting multiple fields from each document
in a dataset, for simplicity of presentation, our formal treatment considers
extracting the value of a single field.

\paragraph{Annotations}
Given a  field $\field$ of a data-set $\dataset$, an \emph{annotation}
$\annotation(\doc)$ of $\doc \in \dataset$ is a list of locations
$[ \location_1, \dots, \location_n ]$ and an aggregation function $\aggregate$
such that $\field(\doc) =
\aggregate(\data[\location_1],\ldots, \data[\location_n])$.
Annotations are user provided ``labels'' in ML parlance, and are used as
training data.
%
%
For our experiments, we built a visual user interface where annotators could
click on individual HTML and image documents to select annotation locations.
In the background, the tool converts these clicks into locations, i.e., XPaths
in HTML and x-y coordinates in PDF documents.

\begin{example}
    \label{ex:annotations}
    For the departure time field in Figure~\ref{fig:formed_documents}(a), the annotation contains the two locations with text elements ``Friday, Apr 3 8:18 PM'' and ``Thursday, Apr 9 2:02 PM'', and the
    aggregation function collects these values into a list.
\end{example}

\paragraph{The formed document extraction problem}
Fix a dataset $\dataset$ and a field $\field$.
The input to the formed document extraction problem is given by
item a set of annotations on a \emph{training set} $\dataset_\train \subseteq
\dataset$.
The ideal expected output for such a problem is an \emph{extraction function}
$\extract$ such that $\forall \doc \in \dataset. \extract(\doc) = \field(\doc)$.
However, it is hard to produce ideal extractions, and we instead use the
standard metrics of \emph{precision}, \emph{recall} and \emph{F1 score} to
measure the quality of an extraction function (see, for example,~\cite{PR}).
In practice, we are usually interested in extracting multiple fields from a
document at once, and in fact, our implementation can do so.
However, for simplicity of discussion, we present our techniques and conduct our
experiments for one field at a time.  

\subsection{Landmarks and Regions}
\paragraph{Landmarks}
A landmark is a value that we can use to identify a location in a
document, such that the field value is present in a ``nearby'' location (or in
"nearby" locations if the field value is aggregated from multiple data values).
Formally, a \emph{landmark} is given by a data value $\landmark$.
A given landmark $\landmark$ identifies a unique location $\location$
in a document $\doc$ such that $\data[\location_i] \supseteq \landmark$, i.e., the
landmark value is a sub-string of the data at $\location_i$.
In order for a landmark to be useful for our purposes, we require the existence
of an inexpensive "locator" function, which can locate the occurrences of a
landmark in a document.
More precisely, we assume a computationally inexpensive function $\locate$ such
that, $\locate(\doc,\landmark) = \location \implies \data[\location] = \landmark$.

\begin{example}
    \label{ex:landmark}
    Consider the travel itinerary document in Figure~\ref{fig:formed_documents}(a).
    In order to extract departure times from this document, a possible landmark
    to use is the phrase ``Depart:''.
\end{example}

\begin{remark}
    For ease of presentation, the definition assumes
    that landmarks occur in one location per document
    (contrast against \emph{Depart} in Figure~\ref{fig:formed_documents}(a)).
    We discuss handling multiple, ambiguous, landmark locations in Section 6.
\end{remark}

\paragraph{Regions}
A {\em region} $\region$ of a document $\doc$ is a set of contiguous locations.
A region can be thought of as a ``sub-document''.
%
%
%
%
Given a set of locations $\locationset$ of a document $\doc$, the \emph{enclosing
region} $\encr(\locationset, \doc)$ is the smallest region that contains all
locations in $\locationset$.
We are particularly interested in regions that enclose a landmark and the
corresponding field values as our approach is based on narrowing down the
document to such regions. 
We call such regions as {\em regions of interest} or {\em ROIs} for short.
    
\begin{example}
    The bottom two blue rectangles in Figure~\ref{fig:formed_documents}(a) highlight the relevant ROIs that contain both the landmark ``Depart:'' and the
    associated field values.
\end{example}

\paragraph{Blueprints}
%
%
Intuitively, a blueprint of a region
is a ``hash'' of all the parts that are ``common'' to
all such regions in the cluster.
%
%
For example, the strings "Airline Record Locator", "AIR", "Meal", "Depart:" in Figure~\ref{fig:formed_documents}(a) are common values, since they will occur in all
documents that follow this format.
%
Let the \emph{layout} of a region $\region$ in a document
$\doc$ be the subset of all locations $\location \in \region$ such that
$\data[\location]$ is a common value. The \emph{blueprint} $\blueprint(\region)$ of a region $\region$ is then defined as a hash
of values in the layout of $\region$.

If two regions are similar, we want to define their blueprints such that they are close to each other.
Given two blueprints $\bpvalue_1$ and $\bpvalue_2$, we use the notation
$\fpd(\bpvalue_1, \bpvalue_2)$ to denote the distance between $\bpvalue_1$ and
$\bpvalue_2$. 
If $\region_1$ and $\region_2$ are similar in structure, we want
$\fpd(\blueprint(\region_1), \blueprint(\region_2))$ to be small value.

\begin{example}
  Our notion of blueprint of an HTML region is based on the XPaths
  to its DOM nodes, but ignoring node order.
  For example, the blueprint of a region stores the path starting from
  a  \texttt{div} node, descending through \texttt{table}, \texttt{tr}, \texttt{td}, and \texttt{span} nodes, but without storing where this path is in relationship to other paths.
\end{example}

\subsection{Landmark-based DSLs}
To formalize the notions of extraction using landmarks, we introduce a special class of DSLs called landmark-based DSLs. This is a generic design of languages that formally captures our reasoning using landmarks and regions for extraction tasks in a domain-agnostic fashion. Figure \ref{fig:dsllandmark} shows the structure of a landmark-based DSL. In such DSLs, the input is always assumed to be a document and a complete program returns a field value extracted from the document. Such DSLs consist of four notions: landmarks $m$, blueprints $b$, region extraction programs $p_{rx}$ and value extraction programs $p_{vx}$.
The region and value extraction programs can be instantiated arbitrarily for a particular domain by defining the language fragments $\mathcal{L}_{rx}$ and $\mathcal{L}_{vx}$, and we shall illustrate such instantiations of these fragments for the web and image extraction domains. 

 These four notions are brought together in the single top-level $\mathsf{Extract}$ operator, the semantics of which is defined in Algorithm  ~\ref{algo:extract}. This operator takes a list $Q$ of 4-tuples, where each tuple consists of a landmark, a region program, a blueprint for the region, and an extraction program. Each tuple represents an extraction strategy for a particular region format. The region program uses a landmark to identify a region of the document, and if this region matches the given blueprint, then the extraction program can be applied on the region to extract the field value. The top-level operator acts as a switch statement that applies the first tuple that successfully extracts a value from the document. Formally, for each tuple, we first use the $\locate$ function to identify the location corresponding to the landmark $\landmark$. This location is then input into the region program $\RProg_i$ to produce the
region of interest $\region$. Now, we proceed with this cluster only if the blueprint of the region is
within a certain tunable  threshold of similarity. If the blueprint is close enough, we return the output of the extraction
program $\EProg_i$ on the region. Otherwise, we continue with the remaining tuples in $Q$.

\begin{figure}
\small{
\begin{tabular}{rcl}
\!\!\!\!$\boldsymbol{\mathsf{@start}}$ \; $\mathsf{T}$ \; $\mathit{t}$ \!\! \!\!\!\!&$:=$&\!\! \!\!\!\!\! $\mathsf{Extract}(\mathit{q,\!...,q})$ where $q = (\mathit{m}, \mathit{p}_{rx}, \mathit{b}, \mathit{p}_{vx}) $ \;\; \\ [0.75ex]
\!\!\!\!$\mathsf{R \!\rightarrow\! T}$ \; $\mathit{p}_{vx}$ \!\! \!\!\!\!&$:=$&\!\! \!\!\!\!\! $\ldots \;\; \LL_{vx} \;\; \ldots$ \;\; \\ [0.75ex]
\!\!\!\!$\mathsf{(doc, str) \!\rightarrow\! R}$ \; $\mathit{p}_{rx}$ \!\! \!\!\!\!&$:=$&\!\! \!\!\!\!\! $\ldots \;\; \LL_{rx} \;\; \ldots$ \;\; \\ [0.75ex]
\!\!\!\!$\mathsf{str}$  $\mathit{m}$ \!\! \!\!\!\!& &\!\! \!\!\!\!\!  \;\; // landmark \\ [0.75ex]
\!\!\!\!$\mathsf{obj}$  $\mathit{b}$ \!\! \!\!\!\!& &\!\! \!\!\!\!\!  \;\; // blueprint \\ [0.75ex]
\!\!\!\!$\boldsymbol{\mathsf{@input}}$ \; $\mathsf{doc}$  $\mathit{d}$ \!\! \!\!\!\!& &\!\! \!\!\!\!\!  \;\; // input document\\ [0.75ex]

\end{tabular}

}
\vspace{-2ex}
\caption{Structure of a Landmark-based DSL $\LL_{ld}$}
\label{fig:dsllandmark}  
\end{figure}

\begin{algorithm}
\small
\caption{Semantics of the $\mathsf{Extract}$ operator in a landmark-based DSL. The blueprint threshold $t$ is a tunable parameter of the semantics.}
\label{algo:extract}
\begin{algorithmic}[1]
  \Require Input document $d$ of type $\doc$
  \Require List $Q = [q_1,...,q_k]$, where each $q_i$  has the form $(m, p_{rx}, b, p_{vx})$ for $ 1 \leq i \leq k$
  \For{$(m, p_{rx}, b, p_{vx}) \in
      Q$}
  \State $\location \gets \locate(d, m)$
  \State $\region \gets p_{rx}(d, \location)$
  \If{$\region \neq \bot \land \fpd(\blueprint(\region), b) \leq t$}
  \State \Return $\aggregate(p_{vx}(\region))$
  \EndIf
  \EndFor
  \State \Return $\bot$
\end{algorithmic}
\end{algorithm}

\section{Landmark and Region based Synthesis}
\label{sec:algorithms}
\label{sec:lrsyn-algo}

In this section we present our generic technique  \technique for synthesizing extraction programs in landmark-based DSLs, and we present instantiations of this technique for different domains in Section~\ref{sec:instantiations}.

\leaveout{
We proceed in a top-down fashion presenting the main algorithm with
Section~\ref{sec:lrsyn-algo} and then detailing two sub-procedures:
\begin{inparaenum}[(a)]
  \item In Section~\ref{subsec:clustering-and-landmark-inference}, we describe 
    how \technique identifies landmarks and clusters an input dataset, and
  \item In Section~\ref{subsec:synth}, we describe the synthesis procedures that synthesize extraction programs in the landmark DSL $\LL_{ld}$.
\end{inparaenum}
}

\subsection{\lrsyn: The Solution Outline}

Our solution data-extraction system is parametrized by a number of components
that need to be instantiated for each domain.
\begin{compactitem}
  \item \emph{Region extraction program synthesizer.}
    The region extraction DSL $\LL_{rx}$ is equipped with a synthesizer that
    takes as input examples of the form $(\doc, \location) \mapsto \region$, and
    produces programs from $\LL_{rx}$.
    Here, the example maps a document $\doc$ and a location $\location$ within
    $\doc$ to a region $\region$ of the document.
    For instance, an example in the HTML domain will have an input HTML
    document, an input location (DOM node), and an output region (set of
    contiguous DOM nodes).
  \item \emph{Value extraction program synthesizer.}
    The value extraction DSL $\LL_{vx}$ is equipped with a synthesizer that
    takes as input examples of the form $\region \mapsto \fvalue$, and produces
    a program from $\LL_{vx}$.
    Here, $\region$ is a region in a document and $\fvalue$ is the field-value
    for that document.
    For instance, an example might have an input region given by the blue
    rectangle in Figure~\ref{fig:formed_documents}(a), and an output value
    ``Friday, Apr 3 8:18 PM''.
  \item \emph{Blueprinting and Locating functions.}
    The blueprinting function $\blueprint$, locating function $\locate$, and the
    blueprint distance function $\fpd$ (as described in
    Section~\ref{sec:problem}) need to be specified per domain.
\end{compactitem}

\begin{figure*}[t]
  \scalebox{0.7}{
\begin{tikzpicture}[scale=0.5, every node/.style={inner sep=5}]
  \node[draw, rectangle] (trainset)
      {\tabular{c}Training Subset\\with Annotations $\annotation$ \endtabular} ;
  \node[rectangle, below of=trainset, yshift=-5mm] (dataset) {Documents} ;
  \node[draw, rectangle, thick, fit=(trainset)(dataset)] (fulldataset) {};

  \node[draw, rectangle, right of=fulldataset, xshift=5cm, yshift=1cm, minimum width=3.7cm] (cl1)
      {\tabular{c}Cluster \& Landmark\\ $(\cluster_1, \landmark_1)$ \endtabular};
  \node[draw, rectangle, dashed, below of=cl1, yshift=-2mm, minimum width=3.7cm] (cl2)
      {$(\cluster_2, \landmark_2)$};
  \node[rectangle, below of=cl2, minimum width=3.7cm, yshift=0.5cm] (cl3)
      {$\ldots$};
  \node[rectangle, below of=cl3, minimum width=3.7cm, yshift=0.5cm] (cl4)
      {$\ldots$};

  \draw[->] (fulldataset.east) -- ++(1.5cm, 0cm) |- node[above] {} (cl1);
  \draw[->, dashed] (fulldataset.east) ++(1.5cm, 0cm) |- node[above] {} (cl2);
  \draw[->, dashed] (fulldataset.east) ++(1.5cm, 0cm) |- node[above] {} (cl3);
  \draw[->, dashed] (fulldataset.east) ++(1.5cm, 0cm) |- node[above] {} (cl4);

  \node[draw, rectangle, right of=cl1, xshift=5cm, yshift=1cm, minimum width=5cm] (inferroi)
    {\tabular{c}For each $\doc_i \in \cluster_1$ \\ Infer ROI $\region_i$ \endtabular};

  \node[draw, rectangle, below of=inferroi, yshift=-0.5cm, minimum width=5cm] (regionsynth)
    {\tabular{c}Synthesize Region Program\\Examples: $(\doc_i, \landmark_i) \mapsto \region_i$ \endtabular} ;

  \node[draw, rectangle, below of=regionsynth, yshift=-0.5cm, minimum width=5cm] (valuesynth)
    {\tabular{c}Synthesize Value Program\\Examples: $\region_i \mapsto \annotation(\doc_i)$ \endtabular} ;

  \node[draw, rectangle, below of=valuesynth, yshift=-0.5cm, minimum width=5cm] (inferbp)
    {\tabular{c}Compute average \\ ROI blueprint \endtabular} ;

  \node[draw, thick, fit=(inferroi)(regionsynth)(valuesynth)(inferbp)] (synth) {};
  \draw[->] (cl1.east) -- (cl1.east-|synth.west) ;

  \node[draw, rectangle, right of=regionsynth, xshift=6cm, yshift=1cm, minimum width=4cm] (rprog)
    {Region Program $\RProg_1$} ;
  \node[draw, rectangle, below of=rprog, minimum width=4cm] (vprog)
    {Value Program $\EProg_1$} ;
  \node[draw, rectangle, below of=vprog, minimum width=4cm] (bpvalue)
    {ROI Blueprint $\bpvalue_1$} ;
  \node[rectangle, above of=rprog, minimum width=4cm,yshift=-0.3cm] (fullprog1)
    {Extraction Program for $\cluster_1$} ;
  \node[draw, thick, fit=(rprog)(vprog)(bpvalue)(fullprog1)] {};

  \draw[->] (regionsynth.east) -- ++(1.00cm,0cm) |- (rprog.west) ;
  \draw[->] (valuesynth.east)  -- ++(1.25cm,0cm) |- (vprog.west) ;
  \draw[->] (inferbp.east)     -- ++(1.50cm,0cm) |- (bpvalue.west) ;

  \node[draw, dashed, below of=fullprog1, yshift=-3cm, inner sep=0.4cm] (fullprog2)
    {Extaction Program for $C_2$};
  \node[below of=fullprog2] (fullprog3) {...};

  \draw [decorate, thick, decoration = {calligraphic brace}]
    ($(cl1.east|-synth.south)+(0cm,-0.5cm)$)
    -- node[below] {\large Joint Cluster and Infer Landmarks}
    ($(fulldataset.west|-synth.south)+(0cm,-0.5cm)$) ;

  \draw [decorate, thick, decoration = {calligraphic brace}]
    ($(fullprog1.east|-synth.south)+(0cm,-0.5cm)$)
    -- node[below] {\large Synthesize Extraction Programs}
    ($(synth.west|-synth.south)+(0cm,-0.5cm)$) ;

\end{tikzpicture}
  }
 \vspace{-2ex}
 \caption{Outline of landmark-based robust synthesis \lrsyn}
 \label{fig:lrsynblocks}
 \vspace{-3ex}
\end{figure*}

\begin{algorithm}
\small
\caption{Landmark-based robust synthesis $\mathsf{LRSyn}$}
\label{algo:lrsyn}
\begin{algorithmic}[1]
  \Require Training set $\dtr \subseteq \dataset$
  \Require Annotation $\annotation(\doc)$ for each document $\doc \in \dtr$
  \Require Region extraction DSL $\LL_{rx}$
  \Require Value extraction DSL $\LL_{vx}$
  \State $[ (\cluster, \landmark) ] \gets \JointClusterAndLandmark(\dtr, \annotation)$
      \label{line:joint-cluster-and-landmark}
  \For{Cluster and landmark $(\cluster_i, \landmark_i) \ \in [(\cluster, \landmark)]$:} 
    \State $(\RProg_i,\bpvalue_i,\EProg_i) \gets$
    \Statex\hspace{\algorithmicindent}$\hspace*{3mm}\SynthesizeExtractionProgram(\cluster_i, \landmark_i, \annotation, \LL_{rx}, \LL_{vx})$
  \EndFor
  \State \Return $\mathsf{Extract}(\{ (\landmark_i, \RProg_i, \bpvalue_i, \EProg_i) \})$
\end{algorithmic}
\end{algorithm}

Algorithm~\ref{algo:lrsyn} presents an outline of our landmark-based robust
synthesis algorithm.
The high-level components are illustrated in Figure~\ref{fig:lrsynblocks}.
Given an annotated training set $\dtr$, the first task is to infer the clustering of
$\dtr$ into $\cluster_0, \ldots, \cluster_n$.
In Algorithm~\ref{algo:lrsyn}, this step is combined with the inference of
landmarks.
The procedure $\JointClusterAndLandmark$
(line~\ref{line:joint-cluster-and-landmark}) produces a set of clusters $C_i$
each associated with a landmark $\landmark_i$.
Then, for each cluster $\cluster_i$, the algorithm calls the subroutine
$\SynthesizeExtractionProgram$ to synthesize a region extraction program $\RProg_i$, a
region blueprint $\bpvalue_i$, and a value extraction program $\EProg_i$.
%
%
The algorithm combines these with the landmark to  output an extraction program in the Landmark-based DSL $\LL_{ld}$, which can be executed using semantics shown in Algorithm 1.

\subsection {Clustering  Documents and Inferring Landmarks}
\label{subsec:clustering-and-landmark-inference}

\leaveout{
In Algorithm~\ref{algo:lrsyn}, the first step is to cluster the set of documents
$\dataset$ into clusters $\{ \cluster_0, \ldots, \cluster_n \}$ and identify a
landmark $\landmark_i$ for each cluster.
Ideally, we want to cluster the documents not based on the global structure, but
the local structure of the ROI around the landmark and field values.
However, the appropriate ROI cannot be identified without the
corresponding landmark, and in turn, landmarks can only be identified given the
clusters as they are defined in terms of values common to all documents in a cluster.
To avoid this circular dependence, we first generate fine-grained clusters based
on the global structure, compute landmarks and regions of interest for these
clusters, and then merge these fine-grained clusters based on the blueprints
of the regions of interest.
}

\begin{algorithm}
\small
\caption{Joint clustering and landmark inference \\
 \textbf{Procedure} $\JointClusterAndLandmark(\dtr,\annotation)$ }
\begin{algorithmic}[1]
\Require Training dataset $\dtr$ along with annotations $\annotation$.
\Require Blueprint function $\blueprint$.
\Require Blueprint distance metric dataset $\fpd$.
%
\State \Comment{\textbf{Initial clustering using whole document blueprints}}
\State $\Delta_{\mathsf{fine}}(\doc, \doc') \gets
        \fpd(\blueprint(\doc), \blueprint(\doc')), \forall \doc, \doc' \in \dtr$
        \label{line:fine-grained-distance}
\State $ [ \cluster ] \gets \mathsf{Cluster}(\dtr, \Delta_\mathsf{fine})$
        \label{line:initial-clustering}
\State \Comment{\textbf{Compute landmark and blueprint candidates}}
\For{$\cluster_i \in [\cluster]$} 
\State $\landmarks_i \gets \LandmarkCandidates(\cluster_i, \annotation)$
    \label{line:landmark-candidates}
\For{$\doc \in \dtr$} \label{line:compute-blueprints-start}
\State $\region_{\doc,\landmark} \gets \encr(\annotation(\doc) \cup
          \locate(\landmark, \doc))$
\Statex\hspace{20em}$, \forall \landmark \in \landmarks_i$
\State $\mathsf{roi}[\doc] \gets \{ (\landmark, \blueprint(\region_{\doc,\landmark})) 
    \mid \landmark \in \landmarks_i \}$
    \label{line:compute-blueprints-end}
\EndFor
\EndFor
\State \Comment{\textbf{Merge clusters}}
\State Define $\Delta_{\mathsf{c}}(\doc_1, \doc_2) \gets$ \par
    $\min(\{ \fpd(\bpvalue_1, \bpvalue_2)
      \mid (\landmark_{1,2}, \bpvalue_{1,2}) \in \mathsf{roi}[\doc_{1,2}] \land
          \landmark_1 = \landmark_2 \}$
    \label{line:coarse-grained-distance}

\While{No change in $[\cluster]$} \label{line:merge-clusters-start}
\State Let $\Delta(\cluster_1, \cluster_2) = 
          \mathsf{Avg}(\{ \Delta_\mathsf{c}(\doc_1, \doc_2) \mid \doc_i \in \cluster_i \})$
\If{$\exists \cluster_1, \cluster_2 \in [\cluster]$ such that $\Delta(\cluster_1, \cluster_2) \leq \mathsf{threshold}$}
  \State $[\cluster] \gets ([\cluster] \setminus \{ \cluster_1, \cluster_2 \})
    \cup \{ \cluster_1 \cup \cluster_2 \}$
    \label{line:merge-clusters-end}
\EndIf
\EndWhile

\State \Return $[\cluster, \mathsf{TopLandmarkCandidate}(\cluster) ]$
\end{algorithmic}
\label{algo:joint-clustering-landmarks}
\end{algorithm}

The \JointClusterAndLandmark\ procedure (Algorithm~\ref{algo:joint-clustering-landmarks}) outlines how we jointly perform clustering and landmark detection, using the approach described in Section~\ref{subsec:joint-infer-cluster}.

\subparagraph{Initial clustering}
Lines~\ref{line:fine-grained-distance}-\ref{line:initial-clustering} perform the initial clustering to obtain the initial fine-grained clusters.
Here, the clustering is by the blueprint of the whole document, and
hence, two documents will be in the same cluster only if they have more or less
exactly the same format with little or no variations.

\subparagraph{Landmark and blueprint identification}
The procedure $\LandmarkCandidates$ identifies common values in the
documents of $\cluster_i$ as landmark candidates and orders
them by a \emph{scoring function} (line~\ref{line:landmark-candidates}).
The scoring function is based on two features:
\begin{inparaenum}[(a)]
  \item the distance between the landmark candidate and the annotated values in
    the document, and
  \item the size of the region that encloses the landmark candidates and
    annotated values.
\end{inparaenum}
These features were determined after initial experiments on a small fraction of
our evaluation datasets.
The procedure $\LandmarkCandidates$ only return candidates with a score
over a certain threshold.
Then, for each landmark candidate and document, we compute and store the
blueprint of the ROI in lines 8-9.

\subparagraph{Coarse-grained clustering}
Now, for each document, we have a number of landmark candidates along with their
associated ROIs.
With the ROIs, we can now define a coarse-grained distance over
documents that is based only on the blueprints of the local structure of ROIs
(line~\ref{line:coarse-grained-distance}).
With this coarse-grained distance, we now repeatedly merge clusters based on
their average document distance
(line~\ref{line:merge-clusters-start}-\ref{line:merge-clusters-end}).
Since the coarse-grained distances are based on the blueprints of ROI, we now
have clusters that are solely based on the local structure,
which was our intention in the first place.

\subparagraph{Finalizing landmarks}
Finally, the procedure returns each coarse-grained cluster along with its top
landmark candidate.


\subsection{Synthesizing Extraction Programs}
\label{subsec:synth}

\begin{algorithm}
\small
\caption{Synthesize Extraction Program\\
  \textbf{Proc.} $\SynthesizeExtractionProgram(
    \cluster,
    \landmark,
    \annotation,
    \LL_{rx},
    \LL_{vx})$}
  \label{algo:synthesize}
\begin{algorithmic}[1]
\Require Cluster $\cluster$ with annotations $\annotation$
\Require Landmark value $\landmark$
\Require Region and value extraction DSLs: $\LL_{rx}$ and $\LL_{vx}$
\State \textbf{for all}~$\doc_i \in \cluster$~\textbf{define}:
\label{line:region-compute-begin}
\State \hspace{\algorithmicindent}
        $\location_i \gets \locate(\landmark, \doc_i)$
\State \hspace{\algorithmicindent}
        $(\locations_i, \aggregate_i) \gets \annotation(\doc_i)$
\State \hspace{\algorithmicindent}
        $\region_i \gets \encr(\{ \location_i \} \cup \locations_i, \doc_i)$
\label{line:region-compute-end}
\State \Comment{\textbf{Synthesize region program}}
\State $\RegionSpec \gets \{ (\doc_i, \location_i) \mapsto \region_i
           \mid \doc_i \in \cluster
      \}$
  \label{line:region-spec}
\State $\RProg \gets \mathsf{Synthesize}(\RegionSpec, \LL_{rx})$
  \label{line:region-synth}
\State \Comment{\textbf{Compute region blueprint}}
\State $\bpvalue \gets
    \mathsf{Average}(\{ \blueprint(\RegionSpec(\doc))) \mid \doc \in \cluster \})$
  \label{line:fp-compute}
\State \Comment{\textbf{Synthesize extraction program}}
\State $\ExtractionSpec \gets \{ \region_i \mapsto
        \aggregate_i(\locations_i) \mid \doc_i \in \cluster \}$
  \label{line:extraction-spec}
\State $\EProg \gets \mathsf{Synthesize}(\ExtractionSpec, \LL_{vx})$
  \label{line:extraction-synth}
\State \Return $(\RProg, \bpvalue, \EProg)$
\end{algorithmic}
\end{algorithm}

The $\SynthesizeExtractionProgram$ procedure (Algorithm~\ref{algo:synthesize}) outlines
how we process a cluster with a given landmark, and calculate a region extraction program, blueprint, and a value extraction program.
The algorithm takes as input:
\begin{inparaenum}[(a)]
  \item a cluster $\cluster$ and corresponding landmark $\landmark$ 
  \item the annotations $\annotation$ for the documents in $\cluster$, and
  \item the DSLs for region programs and extraction programs.
\end{inparaenum}

In the first step, the algorithm computes the ROI $\region_i$
for each document $\doc_i$ from the landmark and the annotations
(lines~\ref{line:region-compute-begin}-\ref{line:region-compute-end}).
Then, we synthesize the region program $\RProg$ using a set of examples of the
form $(\location_i, \doc_i) \mapsto \region_i$
(lines~\ref{line:region-spec} and~\ref{line:region-synth}).
We also compute the average or typical blueprint $\bpvalue$ for all the
ROIs in the cluster (line~\ref{line:fp-compute}).
The region extraction program $\RProg$ and filtering based on blueprint $\bpvalue$ (used in the execution semantics in Algorithm 1) together act as a robust system for detecting the
ROIs.
Next, we synthesize a value extraction program $\EProg$ using examples where the
inputs are the ROIs in the document, and the outputs are the
expected field values (line~\ref{line:extraction-spec}
and~\ref{line:extraction-synth}).
The region and value extraction programs work not only on the documents in $\cluster$, but typically also on unseen documents and formats where the global structure changes, 
without changes in the ROI.

The algorithm finally returns $(\RProg, \bpvalue, \EProg)$, which is combined
with the landmark value $\landmark$ to produce a complete extraction program in the landmark DSL $\LL_{ld}$ in
Algorithm~\ref{algo:lrsyn}.



\section{Instantiating \technique}
\label{sec:instantiations}

%
%

We instantiate $\lrsyn$ for the domains of HTML documents and form images like the 
ones shown in Figure~\ref{fig:formed_documents}, describing in detail the region and value extraction DSLs

\subsection{HTML documents}
\label{sec:html}

%
%
%

\subparagraph{Landmarks and Landmark Candidates}
%
We use \emph{$n$-grams} as landmarks ($n \leq 5$), and the $\locate$ function
lists all the document DOM nodes and finds those containing the landmark
$n$-gram.
The $\LandmarkCandidates$ procedure for identifying the top landmark
candidates lists all $n$-grams in the document, filters out those containing
stop words, retains those $n$-grams common to all documents in the cluster, and
then scores them according to the criteria from Section~\ref{sec:algorithms}.
In particular, the score for a landmark candidate $\landmark$ is given by a
weighted sum of:
\begin{inparaenum}[(a)]
  \item the number of nodes in the path from the DOM nodes corresponding to
    $\landmark$ and field value $\fvalue$, 
  \item the number of nodes in the smallest region enclosing both $\landmark$
    and $\fvalue$, and
  \item the Euclidean distance between $\landmark$ and $\fvalue$ in the rendered
    document.
\end{inparaenum}
%
%
%

\subparagraph{Blueprints}
%
%
We define the blueprint of a region to be the set of XPaths to the \emph{common
value} DOM nodes in the region, ignoring the DOM node order.
For example, the XPath \texttt{body[1]/table[4]/tr[3]/td[2]} is
simplified to \texttt{body/table/tr/td} before adding it to the blueprint set. 

\subparagraph{Region Extraction DSL}
%
%
%
A program in the region extraction DSL $\LL_{rx}$ is a pair of integers $(\parentHops, \siblingHops)$
of \emph{parent hops} and \emph{sibling hops}.
Given a landmark location $\location$, the semantics of the $(\parentHops,
\siblingHops)$ is as follows:
\begin{inparaenum}[(a)]
  \item from $\location$ go up the DOM tree $\parentHops$ steps to get node $n_1$,
  \item from $n_1$ go $\siblingHops$ right to obtain node $n_2$, and
  \item the result is the set of all descendants of all sibling nodes between
    $n_1$ and $n_2$ (inclusive).
\end{inparaenum} 
%
%
For synthesizing program in $\LL_{rg}$ given the landmark location $\location$
and the annotated location $\locations$, we first take the lowest common
ancestor (LCA) $n$ of $\location$ and all nodes in $\locations$.
The $\parentHops$ is given by the difference in depths of $n$ and $\location$
minus $1$, and $\siblingHops$ is given by the difference in index of the
left-most and right-most child of $n$ that have one of $\location$ or
$\locations$ as a descendant.
%
%
%

\subparagraph{Value Extraction DSL}
For the value extraction DSL $\LL_{vx}$, we build upon the synthesis techniques
from~\cite{raza2017automated} and~\cite{flashfill}, as in~\cite{ij2019}.
We do not discuss the DSL and synthesis techniques in detail, but refer the
reader to~\cite{ij2019}.
From a bird's eye view, a program in $\LL_{vx}$ consists of two parts: a web
extraction program which extracts the particular DOM node which contains the
field value, and a text extraction program which extracts the field value from
the text present in the extracted DOM node.
Given an example $\region \mapsto \fvalue$, the synthesis procedure first finds
the DOM node $n$ which contains the text $\fvalue$.
Then, we use $\region \mapsto n$ as the example to synthesize the web extraction
program using techniques from~\cite{raza2017automated}, and $\mathsf{text}(n) \mapsto
\fvalue$ as the example to synthesize the text extraction program using
techniques from~\cite{flashfill}.

\begin{example}
    Consider the task of extracting departure time from the email in Figure~\ref{fig:formed_documents} (a). The synthesized region extraction, value extraction programs and blueprint are shown in Figure~\ref{lst:lrsynprogram}.
\end{example}

\subsection{Form Images}

This domain concerns images that are obtained by scanning or photographing of
physical paper documents.
These images are first processed by an Optical Character Recognition
(OCR) technique to obtain a list of text boxes along with their coordinates.
The form images domain is significantly more complex than the HTML domain as:
\begin{inparaenum}[(a)]
    \item The OCR output is generally very noisy, sometimes splitting up field
      values into a varying number of different text boxes.
    %
    %
    \item These documents do not come equipped with a hierarchical structure
        that defines natural regions.
    %
\end{inparaenum}

\subparagraph{Landmarks and Landmark Candidates}
As in the HTML case, we use $n$-grams as landmarks.
The $\locate$ and $\LandmarkCandidates$ functions work similarly to the HTML
case with OCR output text boxes replacing DOM nodes.
The scoring function for $\LandmarkCandidates$ computes the score for a landmark
candidate $\landmark$ as a weighted sum of:
\begin{inparaenum}[(a)]
  \item the Euclidean distance between $\landmark$ and field value $\fvalue$, and
  \item the area of the smallest rectangle that encloses both $\landmark$ and $\fvalue$.
\end{inparaenum}
%
%

\subparagraph{Blueprints}
Rather than considering all common boxes for blueprinting as in the HTML case,
we instead use only the boxes containing the top $50\%$ most frequent $n$-grams.
The blueprint of a region is defined to be the $\BoxSummary$ of each such box
taken in document order.
The $\BoxSummary$ of  $\textbox$ consists of $2$ parts:
\begin{inparaenum}[(a)]
    \item The frequent $n$-gram that is present in the box, and
    \item For each of the directions top, left, right, and bottom, the content
        type in the text box that immediately neighbors $\textbox$ in the direction.
        The content type of a box is either:
        \begin{inparaenum}[(1)]
            \item $\bot$ if the box does not exist,
            \item the frequent $n$-gram in the text of the box if one exists, and
            \item $\top$ if the box exists, but does not contain a frequent $n$-gram.
        \end{inparaenum}
\end{inparaenum}

\begin{example}
    Consider the text box enclosing \emph{Engine number} in the Accounts Invoice image in
    Figure~\ref{fig:formed_documents}(c).
    The $n$-gram \emph{Engine number} is frequent and hence, is included in the
    blueprint.
    The $\BoxSummary$ of the box is given by:
    $\langle
        \mathsf{ngram} \mapsto \text{\emph{Engine number}},
        \mathsf{Top} \mapsto \bot,
        \mathsf{Left} \mapsto \text{\emph{Chassis number}},$
        $\mathsf{Right} \mapsto \text{\emph{Reg Date}}, 
        \mathsf{Bot} \mapsto \top
    \rangle$
    Here, \emph{Engine number} and \emph{Reg Date} are also a frequent $n$-grams, while the value of
    the Engine number \emph{4713872198212} is not.
\end{example}

\begin{figure}
\begin{alltt}
RProg     := Disjunct(path, path, ...)
path      := input | Expand(path, motion)
motion    := Absolute(dir, k)
           | Relative(dir, pattern, inclusive)
dir       := Top | Left | Right | Bottom
\end{alltt}
\vspace{-2ex}
\caption{The Form Images Region extraction DSL $\LL_{rx}$}
\label{fig:region_dsl}
\end{figure}

\subparagraph{Region Extraction DSL}
%
%
Figure~\ref{fig:region_dsl} depicts a novel region extraction DSL $\LL_{rx}$ for this domain.
$\LL_{rx}$ has the following components:
%
%
      The top operator is a disjunction of \emph{path programs}:
      operationally, these programs are executed in sequence and the first
      non-null result is returned.
      Due to the OCR noise and variations in form images, often a single
      non-disjunctive path program is not sufficient.
      %
      %
%
      Each path program starts at the input landmark and repeatedly extends the
      path in steps till the path's bounding box covers all the annotated
      values.
      Each extension step is specified by a direction and a motion.
      The motion may be \emph{absolute} (e.g., move right by $4$ boxes) or
      \emph{relative} (e.g., move down till you hit a text box that matches
      the regex \texttt{[0-9]\{5\}}).
      The additional $\mathtt{inclusive}$ parameter indicates whether the box
      that matches the pattern should be included or excluded in the path.

\begin{example}
    \label{ex:pathexample}
    In Figure~\ref{fig:formed_documents}(c), let us consider the landmark \emph{Chassis number} and the
    annotated value \emph{WDX 28298 2L SHX 3 }.
    The field value here is a variable-length string and the OCR splits the
    value into $1-4$ separate boxes.
    Consider the two region programs given below:\\
    $\bullet~\mathsf{Ext(Ext(input, Abs(down, 1)), Rel(Right, [0-9]\{13\}, false))}$\\
    $\bullet~\mathsf{Ext(Ext(input, Abs(down, 1)), Rel(Right, DATE, false))}$\\
    Both programs first move one step down from the landmark \emph{Chassis
    number}.
    However, the first moves to the right till it hits a $13$ digit engine
    number, while the other till it hits a date.
    In case the engine number is present in a given form, the first program
    produces a path which ends with the annotated value, while the second one
    does so if the engine number is absent.
    When combined disjunctively in the right order, they together cover both
    cases.
    %
    %
\end{example}

The synthesis algorithm for $\LL_{rx}$ is split into two parts: generating path
programs and selecting path programs to construct a disjunction.
%
%
Fix a set of input documents with annotations.
We first synthesize path programs for small subsets (size $\leq 3$) of input
documents.
%
For synthesizing path programs, we use \emph{enumerative
synthesis}~\cite{transit,sygus} to generate numerous candidate programs and then
filter them by whether they cover the annotated values when starting from the
landmark.
%
We enumerate paths of up to $4$ motions, bounding $\mathtt{k}$ to positive
integers $< 5$.
For $\mathtt{pattern}$, we enumerate a finite set of regular expression patterns
generated using a string profiling technique \cite{flashprofile,confminer}
over all the common and field text values present in the cluster.
%
%
For example, when given a cluster of documents of similar to
Figure~\ref{fig:formed_documents}c), one of the patterns returned is $\mathtt{[0-9]\{13\}}$ as
the cluster contains many engine numbers of that form. 

%
%
%
%

After the enumeration step, we have a collection $\{ P_1, \ldots, P_n \}$ of
path programs that are each synthesized from a small subset of input examples.
Now, we use the $\ndsyn$ algorithm from~\cite{ij2019} to select a subset of these
programs to construct the disjunctive program. 
Now, for each program $P_i$, we define the set $\mathsf{Ex}_i$ to be the subset of
$\mathsf{Examples}$ that $P_i$ is correct on.
The $\ndsyn$ algorithm selects a subset of $\mathcal{P}$ of programs such that
$\bigcup_{P_i \in \mathcal{P}} \mathsf{Ex}_i = \mathsf{Examples}$, optimizing
for F1 score and program size~\cite{ij2019}.

\begin{example}
   Consider the two path programs from Example~\ref{ex:pathexample}, along with the
   additional program $\mathsf{Ext(Ext(input, Abs(Down, 1)), Abs(Right, 2))}$.
   Given a collection of such path programs, $\ndsyn$ builds the disjunctive
   program using the two from Example~\ref{ex:pathexample} as they cover a large
   fraction of the documents in the cluster.
   The additional program above will be ignored as it is only correct when the
   chassis number field value is split into $2$ boxes by the OCR.
\end{example}

\subparagraph{Value Extraction DSL}
For the value extraction DSL, we use FlashFill~\cite{flashfill}.
The input to the value extraction program is the concatenation of all the
text values in the boxes returned by the path program.

\section{Discussion}

\subsection{Hierarchical Landmarks}

Consider again the email in Figure~\ref{fig:formed_documents}b) and a variant where the term
\emph{Pick-up} has been replaced by \emph{Depart}.
Now, using the landmark \emph{Depart} for extraction will unintentionally also extract the car trip departure time.
%
In Section~\ref{sec:overview}, the human used a hierarchy of landmarks (i.e.,
\emph{AIR} followed by \emph{Depart}) to obtain the correct results.
%
%
%
Algorithm~\ref{algo:lrsyn} can similarly be extended to hierarchical
extractions.

%
%
Say we first synthesized a program $\FullProg_0$ using
Algorithm~\ref{algo:lrsyn} that uses the landmark \emph{Depart}.
%
%
We run $\FullProg_0$ on the training document and realize that a spurious
landmark location (car departure) is identified.
%
%
In this case, we use the correct landmark locations (i.e., the first and last
occurrence of \emph{Depart}) as a new annotation.
Running Algorithm~\ref{algo:lrsyn} with this annotation will produce a program
$\FullProg_1$ that uses \emph{AIR} as a landmark and extracts precisely the
relevant occurrences of \emph{Depart}.
At inference time, we run $\FullProg_1$ to identify only the correct occurrences
of \emph{Depart} and then run $\FullProg_0$ starting with only those
occurrences of \emph{Depart}.
In our implementation in Section~\ref{sec:evaluation}, we have implemented the
full hierarchical extraction algorithm for the HTML domain.

\subsection{Robustness of \technique}

By design, \technique is robust to format variations that change
\begin{inparaenum}[(a)]
  \item document structure outside the ROIs,
  \item position of ROIs, and
  \item the number of ROIs.
\end{inparaenum}
However, there are $2$ clear limitations to the robustness of \lrsyn:
%
  If a format changes by adding a new part (e.g., car departure time) that
    contains the landmark the \lrsyn generated program uses, the program may
    generate a spurious output.
    This is only a problem if the new part added also has similar blueprint to
    the existing ROIs.
    For example, the program will not be mislead by an new banner advertisement
    saying \emph{Depart today for your dream destination!}
    %
  %
  The second case is when the format inside the ROI changes.
    In this case, the underlying assumption about invariant local structure
    is violated and \lrsyn is unlikely to cope with this variation.
    One possible solution that we discuss in Section~\ref{sec:conclusion} is to use a trained ML model to automatically re-synthesize as  in~\cite{ij2019}.

\section{Evaluation}
\label{sec:evaluation}
We evaluate \technique in two scenarios, each with different document types:
(1) HTML documents from travel reservation emails and
(2) Form image documents from invoices, receipts, and travel reservation emails.
In each case, we perform experiments in $2$ settings: \emph{contemporary} and \emph{longitudinal}.
In the contemporary setting, the training data and test data consist of
documents authored and collected during the same time period.
On the other hand, for the longitudinal setting, the test data was collected
several months after the training data, allowing for new formats to organically
enter the dataset.
For both domains, we are attempting to answer the following research question:

\begin{mdframed}[linecolor=white]
\vspace{1ex}
\centering
\emph{How does $\resyn$ compare with previous approaches in the contemporary and longitudinal settings?}
\vspace{1ex}
\end{mdframed}

Before describing the results, we first discuss the $3$ threshold parameters
that control various aspects of learning and inference in \lrsyn.
We pick both the cluster merging threshold in
Algorithm~\ref{algo:joint-clustering-landmarks} and the blueprint distance
threshold in Algorithm~\ref{algo:extract} to be $0$, i.e., an \emph{exact match}.
Note that the blueprint functions already lose information--hence, an exact
match does not mean the ROIs need to have exactly the same format.
For the score threshold for the procedure $\LandmarkCandidates$, we pick a
threshold that gives us around $10$ landmark candidates in each case.
%

We discuss the results on the HTML and form image documents in
Sections~\ref{sec:html-eval} and~\ref{sec:forms-eval}, and discuss some secondary
results on the nature of programs learned by $\lrsyn$ in
Section~\ref{sec:programs-eval}.
In Section~\ref{sec:results-robustness}, we discuss how robust the experimental
results are to various choices in the experimental setup.

\subsection{HTML extraction}
\label{sec:html-eval}
Our HTML document dataset, called the {\em machine-to-human (M2H)} email
dataset, consists of anonymized flight reservation emails.
It consists of $3503$ emails from $6$ different flight providers and is divided
into training and test sets of size $362$ and $3141$, respectively.
For each provider, we synthesize and test programs using Algorithm~\ref{algo:lrsyn},
as instantiated in Section~\ref{sec:html}.
We compare against $2$ state-of-the-art techniques, namely
$\ndsyn$~\cite{ij2019} and $\textsf{ForgivingXPaths}$~\cite{omari2017synthesis}.

\paragraph{Overall results}
Table~\ref{table:m2hoverall}~shows the average precision, recall and F1 scores
across various extraction tasks for $\fx$, $\ndsyn$ and $\lrsyn$ for both
contemporary and longitudinal setting.
As seen in the table, $\lrsyn$ has near perfect precision and recall, with
$\ndsyn$ performing quite well with numbers $>0.9$.
Further, as expected, the gap between $\lrsyn$ and $\ndsyn$ is higher in the longitudinal dataset indicating that $\lrsyn$ can cope with format variations better than $\ndsyn$.

Unlike $\ndsyn$ or $\lrsyn$ which use a combination of structure and text
programs for extraction, $\textsf{ForgivingXPaths}$ only outputs XPaths which
correspond to the entire node, rather than the sub-text contained within that
node.
Consequently, it has high recall and poor precision when the field value is a substring of the entire DOM node text.
We therefore omit it from the more detailed results below.

\begin{table}
\scriptsize
\begin{tabular}{|l|c|c|c|c|}
\hline
\multicolumn{4}{|c|}{Contemporary} \\
\hline
Metric & $\fx$ & $\ndsyn$ & $\lrsyn$\\
\hline
Avg. Precision & 0.17 & 0.96 & 1.00\\ \hline
Avg. Recall & 0.99 & 0.91 & 1.00\\ \hline
Avg. F1 & 0.22 & 0.93 & 1.00 \\ \hline
\end{tabular}
\begin{tabular}{|l|c|c|c|c|}
\hline
\multicolumn{4}{|c|}{Longitudinal} \\
\hline
Metric & $\fx$ & $\ndsyn$ & $\lrsyn$\\
\hline
Avg. Precision & 0.15 & 0.99 & 1.00\\ \hline
Avg. Recall & 0.98 & 0.89 & 1.00\\ \hline
Avg. F1 & 0.20 & 0.92 & 1.00 \\ \hline
\end{tabular}
\caption{Overall scores of $\lrsyn$ and $\ndsyn$ on the M2H Contemporary and Longitudinal datasets}
\label{table:m2hoverall}
\end{table}

\paragraph{Detailed comparison}
Table~\ref{tab:m2hexpmt1} shows a more detailed drill-down of the F1 scores for
$\ndsyn$ and $\resyn$, in the two settings.
%
%
In summary, in all cases, the hit to the F1-scores come from lower precision numbers
rather than lower recall numbers.

\highlightbox{
$\lrsyn$ is very robust to variations in the longitudinal
setting, achieving $>95\%$ F1 score in all $53$ out of $53$ fields, with a perfect F1 score of $1.00$ in $49$ cases.
In comparison, $\ndsyn$ achieves $>95\%$ and perfect scores in $40$ and $33$ cases respectively.
}
In many applications, having a score of $1.00$ is crucial, i.e., even a system
with $0.99$ precision cannot be deployed in practice.
For example, even a tiny imprecision in a system that automatically adds
calendar entries based on flight reservation emails is disastrous for millions
of users.
Comparing the numbers precisely, $\lrsyn$ outperforms $\ndsyn$ in 19 and 20 out
of the 53 fields in the contemporary and longitudinal setting, respectively.
In the remaining fields, the two approaches have comparable F1 scores.

We examined the domains \emph{aeromexico} and \emph{mytrips.amexgbt} where both
$\lrsyn$ and $\ndsyn$ achieved perfect scores. 
In the \emph{aeromexico} domain, each field has a unique dedicated ID attribute
in the HTML domain which act as \emph{implicit landmarks}, and both $\ndsyn$ and
$\lrsyn$ are able to latch on to this unique ID.
For example, the arrival time and departure city DOM nodes have id
\emph{arrival-city} and \emph{departure-city}, and $\ndsyn$ produces a program
that searches for this ID across the whole document, emulating the landmark
location step of a $\lrsyn$ program.

In the \emph{mytrips.amexgbt} domain, the $\ndsyn$ program while perfectly
accurate on all the variations in our dataset, is very fragile.
In the final CSS selector step of the web extraction component, it looks for the $10^{th}$ child of a DOM element corresponding to a flight details section.
Incidentally, all new sections (car reservations, hotel reservations, etc) added
in the new variations have at most $5$ children, and hence, they are
automatically ignored by the $\ndsyn$ program.
Any new variation that would add a long enough section will break this program.
In contrast, the $\lrsyn$ program narrows down on the right region with a
landmark and is resistant to such variations.

\begin{table*}
\scriptsize
\begin{tabular}{|l|c|c|c||c|c|}
\hline
 & & \multicolumn{2}{|c||}{ Contemporary} &  \multicolumn{2}{|c|}{Longitudinal} \\
 {Fields} & Domain & $\ndsyn$ & $\resyn$ & $\ndsyn$ & $\resyn$ \\
\hline
\hline
 AIata & \multirow[b]{4}{*}{ifly}   & 0.81 & \textbf{1.00} & 0.64 & \textbf{1.00} \\
 ATime &                            & 0.76 & \textbf{1.00} & 0.62 & \textbf{1.00} \\
 DIata &                            & 0.73 & \textbf{1.00} & 0.55 & \textbf{1.00} \\
 DDate &                            & 1.00 &         1.00  & 1.00 &         1.00  \\
 DTime &  alaska                    & 0.73 & \textbf{1.00} & 0.55 & \textbf{1.00} \\
 FNum  & \multirow[t]{4}{*}{air}    & 1.00 &         1.00  & 1.00 &         1.00  \\
 Name  &                            & 1.00 &         1.00  & 0.99 &         0.99  \\
 Pvdr  &                            & --   &         --    & --   &         --    \\
 RId   &                            & 1.00 &         1.00  & 1.00 &         1.00  \\

\hline
 AIata & \multirow{9}{*}{airasia} & 0.67 & \textbf{1.00} & 0.67 & \textbf{1.00} \\
 ATime &                          & NaN  & \textbf{1.00} & NaN  & \textbf{1.00} \\
 DIata &                          & 0.67 & \textbf{1.00} & 0.67 & \textbf{1.00} \\
 DDate &                          & 0.67 & \textbf{1.00} & 0.67 & \textbf{1.00} \\
 DTime &                          & NaN  & \textbf{1.00} & NaN  & \textbf{1.00} \\
 FNum  &                          & 1.00 &          1.00 & 0.96 & 0.96          \\
 Name  &                          & 1.00 &          1.00 & 1.00  & 1.00 \\
 Pvdr  &                          & 1.00 &          1.00 & 0.96 & 0.96          \\
 RId   &                          & 1.00 &          1.00 & 1.00 & 1.00          \\

\hline 
\end{tabular}
\begin{tabular}{|c|c|c||c|c|}
\hline
 & \multicolumn{2}{|c||}{ Contemporary} &  \multicolumn{2}{|c|}{Longitudinal} \\
 {Domain} & $\ndsyn$ & $\resyn$ & $\ndsyn$ & $\resyn$ \\
\hline
\hline\multirow{9}{*}{getthere}

 & 0.75 & \textbf{1.00} & 0.74 & \textbf{1.00} \\
 & 0.94 & \textbf{1.00} & 0.91 & \textbf{1.00} \\
 & 0.94 & \textbf{1.00} & 0.95 & \textbf{1.00} \\
 & 0.98 & \textbf{1.00} & 0.95 & \textbf{1.00} \\
 & 0.76 & \textbf{1.00} & 0.78 & \textbf{1.00} \\
 & 0.98 & \textbf{1.00} & 0.98 & \textbf{1.00} \\
 & 1.00 & 1.00 & 0.89 & \textbf{1.00} \\
 & 0.98 & \textbf{1.00} & 0.97 & \textbf{1.00} \\
 & 0.93 & \textbf{1.00} & 0.94 & \textbf{1.00} \\

\hline\multirow{9}{*}{delta}
 & 1.00 & 1.00 & 1.00 & 1.00 \\
 & 1.00 & 1.00 & 1.00 & 1.00 \\
 & 1.00 & 1.00 & 1.00 & 1.00 \\
 & 0.94 & \textbf{1.00} & 0.95 & \textbf{1.00} \\
 & 1.00 & 1.00 & 1.00 & 1.00 \\
 & 1.00 & 1.00 & 1.00 & 1.00 \\
 & 0.85 & \textbf{0.97} & 0.91 & \textbf{0.97} \\
 & 1.00 & 1.00 & 1.00 & 1.00 \\
 & 1.00 & 1.00 & 1.00 & 1.00 \\

\hline 
\end{tabular}
\begin{tabular}{|c|c|c||c|c|}
\hline
 & \multicolumn{2}{|c||}{ Contemporary} &  \multicolumn{2}{|c|}{Longitudinal} \\
 {Domain} & $\ndsyn$ & $\resyn$ & $\ndsyn$ & $\resyn$ \\
\hline
\hline\multirow[b]{4}{*}{aero} & 1.00 & 1.00 & 1.00 & 1.00 \\
                               & 1.00 & 1.00 & 1.00 & 1.00 \\
                               & 1.00 & 1.00 & 1.00 & 1.00 \\
                               & 1.00 & 1.00 & 1.00 & 1.00 \\
\multirow[t]{4}{*}{mexico}     & 1.00 & 1.00 & 1.00 & 1.00 \\
                               & 1.00 & 1.00 & 1.00 & 1.00 \\
                               & 1.00 & 1.00 & 1.00 & 1.00 \\
                               & 1.00 & 1.00 & 1.00 & 1.00 \\
                               & 1.00 & 1.00 & 1.00 & 1.00 \\

\hline
\multirow[b]{4}{*}{mytrips} & 1.00 & 1.00 & 1.00 & 1.00 \\
                            & 1.00 & 1.00 & 1.00 & 1.00 \\
                            & 1.00 & 1.00 & 1.00 & 1.00 \\
                            & 1.00 & 1.00 & 1.00 & 1.00 \\
 amex                          & 1.00 & 1.00 & 1.00 & 1.00 \\
 \multirow[t]{3}{*}{gbt}   & 1.00 & 1.00 & 1.00 & 1.00 \\
                            & 1.00 & 1.00 & 1.00 & 1.00 \\
                            & 1.00 & 1.00 & 1.00 & 1.00 \\
                            & 1.00 & 1.00 & 1.00 & 1.00 \\

\hline 
\end{tabular}
\caption{F1 scores of \ndsyn\ and \lrsyn\ for M2H HTML dataset. The Pvdr field is not relevant for iflyalaskaair}
\label{tab:m2hexpmt1}
\vspace{-5ex}
\end{table*}

\begin{table}
\scriptsize
\begin{tabular}{|c|l|c|c|c|}
\hline
{Domain} & Fields & $\afr$ & $\resyn$\\
\hline\multirow{9}{*}{AccountsInvoice}
  & Amount & 0.99 & \textbf{1.00} \\
  & Chassis & 0.82 & \textbf{0.99} \\
  & CustAddr & 0.98 & \emph{0.96} \\
  & Date & 0.93 & \textbf{0.98} \\
  & Dnum & 0.96 & \textbf{0.97} \\
  & Engine & 0.82 & \textbf{1.00} \\
  & InvoiceAddress & 0.90 & \textbf{0.95} \\
  & Model & 0.75 & \textbf{1.00} \\

\hline\multirow{8}{*}{CashInvoice}
  & Amount & 1.00 & 1.00 \\
  & Chassis & 0.99 & 0.99 \\
  & CustAddr & 0.99 & \emph{0.97} \\
  & Date & 0.99 & 0.99 \\
  & Dnum & 0.96 & 0.96 \\
  & Engine & 0.93 & \textbf{0.95} \\
  & InvoiceAddress & 0.99 & 0.99 \\
  & Model & 0.99 & \textbf{1.00} \\

\hline\multirow{5}{*}{CreditNote}
  & Amount & 1.00 & 1.00 \\
  & CreditNoteAddress & 0.99 & \textbf{1.00} \\
  & CreditNoteNo & 0.94 & \emph{0.93} \\
  & CustRefNo & 1.00 & 1.00 \\
  & Date & 1.00 & 1.00 \\
  & RefNo & 1.00 & 1.00 \\

\hline\multirow{6}{*}{SalesInvoice}
  & Amount & 1.00 & 1.00 \\
  & CustomerReferenceNo & 1.00 & 1.00 \\
  & Date & 1.00 & 1.00 \\
  & InvoiceAddress & 0.94 & \textbf{0.99} \\
  & RefNo & 0.99 & 0.99 \\
  & SalesInvoiceNo & 0.99 & 0.99 \\

\hline\multirow{6}{*}{SelfBilledCreditNote}
  & Amount & 1.00 & 1.00 \\
  & CustomerAddress & 1.00 & \emph{0.99} \\
  & CustomerReferenceNo & 0.99 & 0.99 \\
  & Date & 1.00 & 1.00 \\
  & DocumentNumber & 1.00 & 1.00 \\
  & VatRegNo & 1.00 & 1.00 \\

\hline
\end{tabular}
\caption{F1 scores for Finance dataset}
\label{tab:financedataset}
\vspace{-3ex}
\end{table}

\begin{table}
\scriptsize
\begin{tabular}{|l|p{1.5cm}|c|c|}
\hline
Fields & {Domain} & $\afr$ & $\resyn$\\
 \hline
AIata & \multirow{9}{*}{aeromexico}        & 0.62 & \textbf{0.65} \\
ATime &                                 & 0.69 & \textbf{0.99} \\
DIata &                                 & 0.36 & \textbf{0.66} \\
DDate &                                 & 0.71 & \textbf{0.89} \\
DTime &                                 & 0.65 & \textbf{0.97} \\
FNum  &                                 & 0.66 & \textbf{0.83}  \\
Name  &                                 & 0.96 & \textbf{0.98}  \\
Pvdr  &                                 & 0.69 & \textbf{0.78}  \\
RId   &                                 & 1.00 & 1.00            \\
 
\hline
\end{tabular}
\begin{tabular}{|p{1.5cm}|c|c|}
\hline
{Domain} & $\afr$ & $\resyn$\\
\hline
\multirow{9}{*}{getthere}  & 0.94 & \textbf{1.00} \\
                           & 0.87 & \textbf{1.00} \\
                           & 0.93 & \textbf{1.00} \\
                           & 0.96 & \textbf{0.99} \\
                           & 0.88 & \textbf{1.00} \\
                           & 0.94 & \textbf{1.00} \\
                           & 0.99 &         0.99  \\
                           & 0.75 & \textbf{1.00} \\
                           & 0.89 & \textbf{0.95} \\
 
\hline
\end{tabular}

\begin{tabular}{|l|p{1.5cm}|c|c|}
\hline
{Fields} & Domain & $\afr$ & $\resyn$\\
\hline
AIata & \multirow[b]{4}{*}{ifly.}     & 0.99 & \textbf{1.00} \\
ATime &                                    & 0.95 &   \textbf{1.00} \\
DIata &                                    & 0.98 & \textbf{1.00} \\
DDate &                                    & 0.98 & \emph{-} \\
DTime &  \multirow[t]{4}{*}{alaskaair}       & 0.95 & \textbf{0.98} \\
FNum  &                                    & 0.97 & \textbf{1.00} \\
Name  &                                    & 0.98 & 0.98 \\
Pvdr  &                                    & 0.93 & \textbf{0.99} \\
RId   &                                    & 1.00 &         \emph{0.86} \\
 
\hline
\end{tabular}
\begin{tabular}{|p{1.5cm}|c|c|}
\hline
{Domain} & $\afr$ & $\resyn$\\
 \hline
\multirow[b]{4}{*}{mytrips}  & 0.85 & \textbf{0.98} \\
                             & 0.97 & \textbf{1.00} \\
                             & 0.96 & \textbf{0.99} \\
                             & 0.93 & \textbf{1.00} \\
\multirow[t]{4}{*}{amexgbt}  & 0.99 & \textbf{1.00} \\
                             & 0.98 & \textbf{1.00} \\
                             & 0.98 & \textbf{1.00} \\
                             & 0.91 & \textbf{0.99} \\
                             & 0.61 & \textbf{0.96} \\

\hline
\end{tabular}
\caption{F1 Score for M2H-Images Dataset}
\label{tab:m2himages}
\vspace{-2ex}
\end{table}

\subsection{Form Image Extraction}
\label{sec:forms-eval}

We consider two datasets of form images:
\begin{compactitem}
\item {\em Finance} dataset:
This consists of ~850 images of receipts, purchase
orders, credit notes, sales invoice and similar such documents.
Here, the training and test data are from the same time period and we only
evaluate the contemporary setting.
\item {\em M2H-Images} dataset:
We convert M2H emails from $4$ domains to images and extract the same fields as
before.
This represents common scenarios in practice where HTML documents such
as booking mails or receipts may be printed and then scanned again, say when
expense reports are filed.
The OCR service we use produced extremely poor
results on $2$ of the $6$ domains from the HTML experiments, and hence, we used only $4$ domains in this dataset (The same OCR service is used by our baseline as well, see below).
\end{compactitem}
For both datasets, we use a training data size of $10$ for each field, and compare
against Azure Form Recognizer (\afr)~\cite{AFR}, a cloud-based form data extraction service.

\paragraph{Overall Results}
Table~\ref{tab:imagesoverall}~shows the average precision, recall, F1 and
accuracy scores for \afr\ and \ndsyn\, for both the Finance and M2H-image
datasets.
As we can see from the table, both $\lrsyn$ and $\afr$ perform very well on the
Finance dataset, with $\lrsyn$ performing marginally better.
In this dataset, the image formats do not vary much, resulting in these
high-quality results.
\highlightbox{
$\lrsyn$ outperforms $\afr$, a state-of-the-art industrial neural form extraction
system with just $10$ training images per
field, having a precision of $0.97$ vs $0.90$ on the M2H-Images dataset.
}

\begin{table}
\scriptsize
\parbox{.45\linewidth}{
\centering
\begin{tabular}{|l|c|c|c|}
\hline
Metric & $\afr $ & $\lrsyn$\\
\hline
Avg. Pre.  & 0.98	& 0.99 \\ \hline
Avg. Rec.  & 0.96	& 0.99 \\ \hline
Avg. F1  & 0.97	& 0.99 \\ \hline
\end{tabular}
\caption*{Finance dataset}
}
\hfill
\parbox{.45\linewidth}{
\centering
\begin{tabular}{|l|c|c|c|}
\hline
Metric & $\afr $ & $\lrsyn$\\
\hline
Avg. Prec.  & 0.90	& 0.97 \\ \hline
Avg. Rec.  & 0.93	& 0.97 \\ \hline
Avg. F1  & 0.91	& 0.97 \\ \hline
\end{tabular}
\caption*{M2H-Images dataset}
}
\vspace{-2ex}
\caption{Average precision, recall, F1 numbers on Finance and M2H-Images dataset (Ignoring DDate field in ifly.alaskaair)}
\label{tab:imagesoverall}
\vspace{-3ex}
\end{table}

\paragraph{Detailed comparison}
Table~\ref{tab:financedataset}~shows the results of $\afr$ and $\lrsyn$ on the Finance dataset with respect to $34$ extraction tasks.
$\lrsyn$ performs better than $\afr$ in $12$ out of the $34$ cases and is on par on the rest, with significant gains in some domains like "AccountsInvoice". 

Though the neural model in $\afr$ is trained with thousands of invoices and
receipts, and further fine-tuned with our training data, we observe that it is
sensitive to the region coordinates in a given document.
If these regions are translated, or if the document scan is tilted, $\afr$
produces erroneous results.
On the other hand, $\lrsyn$ is partially robust to such changes as we use a text
landmark.
$\afr$ is marginally better than $\lrsyn$ in some extraction tasks.
These are cases where there is no clear bounding pattern for the field values.
On the other hand, $\afr$'s semantic understanding of the data is not
affected by boundary text patterns.

Table~\ref{tab:m2himages}~shows the results of $\afr$ and $\lrsyn$ on the M2H
dataset with respect to $45$ extraction tasks.
This dataset exhibits more variations at the visual level as compared to the
Finance dataset, and hence, $\lrsyn$ performs better than $\ndsyn$ in $35$ out
of the $45$ tasks and is on par on most of the remaining extraction tasks.
There is $1$ specific case where $\lrsyn$ fails altogether, producing no
programs.
These are cases where there is no local textual landmark geometrically near the
field value.
However, the region around the field value may still be similar across documents.
We discuss the possibility of using visual landmarks as opposed to textual
ones in Section~\ref{sec:conclusion}.

\paragraph{Summary of results}
In the HTML domain, the prior work \ndsyn\ is a high-performing system with F1 scores in the range of 0.9. \lrsyn\ is able to push the F1 scores to a perfect 1.0 in most cases. In longitudinal scenarios, \lrsyn\ improves \ndsyn\ in 20 out of 53 fields, with significant lift in F1 scores in many cases.
In the images domain, even with very little training data, \lrsyn\ matches AFR, which is a released product on contemporary settings, and outperforms AFR in longitudinal settings. In addition, \lrsyn\ produces simpler interpretable programs that match human intuition for extraction, and are much easier to maintain. Hence, we see a lot of promise in this approach.

\subsection{Nature of Synthesized Programs}
\label{sec:programs-eval}

Additionally, we performed secondary
analysis to understand the features of the $\lrsyn$ programs.

\paragraph{Program size}
In the HTML domain, we compared size of the $\lrsyn$ and $\ndsyn$ programs.
Since the programs are naturally of different shapes, we only compare the web extraction part of the programs.
The final text extraction program is generally the same across both algorithms.
Note that these numbers need to be taken in the context that $\lrsyn$ programs
additionally have a landmark and blueprint.
\highlightbox{
  For the M2H dataset, the web extraction part of $\lrsyn$ programs have $2.95$
  CSS selector components as compared to $8.51$ for $\ndsyn$.
}
\paragraph{Quality of Inferred Landmarks}
We infer landmarks automatically using the techniques and scoring functions from Sections~\ref{sec:algorithms} and~\ref{sec:instantiations}.
To check the quality of landmark inference, we also asked data annotators to tag landmarks manually.
\highlightbox{
In $57$ out of $63$ clusters across all fields, the inferred landmarks are
the same as manually provided landmarks.
In $5$ of the remaining $6$ cases, the human annotator agreed that the inferred landmark 
was of equal quality.
}
In the remaining $1$ case, the algorithm chose the human annotated landmark as well, but 
in addition chose a disambiguating hierarchical landmark of low quality.
In particular, it disambiguated the term \emph{Name} occurred in reference to both
the name of the passenger and the name in the  billing address.
Here, the algorithm chose to disambiguate using the term \emph{Meal}, i.e.,
passengers have a meal preference while the billed person does not.
%

\subsection{Robustness of Experimental Results}
\label{sec:results-robustness}


\paragraph{Training set choice}
In our experiments, the training set is small compared to the
full dataset leading to a possibility of over-fitting, with different training
sets potentially producing significantly differing results.
However, our techniques are robust even with small training sets:
\begin{inparaenum}[(a)]
  \item Landmark identification can leverage the full dataset of both labeled
    and unlabeled documents.
  \item $\lrsyn$ does not need to see all format variations that differ only
    outside the ROIs, and a small set covering only the variations within the ROIs is sufficient.
\end{inparaenum}
To confirm this, we reran all our experiments on the M2H-HTML dataset with $4$
different randomly chosen training datasets.
In all runs, the F1 scores of the generated programs for each field and domain
varied by no more than $0.01$ from the results presented in
Table~\ref{tab:m2hexpmt1}, confirming our hypothesis that the results are robust
to training set choice.

\paragraph{Landmark identification threshold}
%
%
For the landmark candidate score threshold, we picked threshold values that
resulted in \textasciitilde$10$ candidates for each case.
To study the robustness of the results to this choice, we reran all experiments
with a threshold value that returned $2X$ as many candidates.
The obtained results were exactly identical to the results presented in the previous sections. 
This is expected as ``bad'' landmark candidates are eliminated in subsequent
steps, i.e., there is usually no program that extracts the required field value
starting from the landmark.
Hence, as long as the threshold is high enough to allow for some good landmark
candidates, it does not matter how many bad landmark candidates are included.

\section{Related work}
\paragraph{Program synthesis}
Data extraction has been an active area of investigation in the program synthesis community.
FlashExtract \cite{lg2014} synthesizes programs from examples for extraction from large text or web documents, using an algebra of pre-defined operators such as map, reduce and filter. FlashExtract works well when inputs are homogeneous. For heterogeneous inputs,  several works have explored \emph{disjunctive} synthesis approaches \cite{alur2015, alur2017, reynolds2015, saha2015, raza2018}. \emph{Forgiving XPaths} ~\cite{omari2017synthesis}~ in particular focuses on synthesizing progressively relaxed XPaths to increase recall for web extraction. Hybrid synthesis ~\cite{raza2020web} proposes a combination of deductive \cite{pg2015} and predictive ~\cite{raza2017automated}~ synthesis techniques to create programs that follow aligned structures inferred on webpages, but this is primarily focused on tabular data extraction. Raychev et al. \cite{raychev2016} also learn programs from noisy data, but generate a single program from a
noisy dataset rather than using clustering to learn different programs applicable to different region formats. In contrast to the above techniques, we propose the idea of landmarks and regions as a compositional approach that addresses the high noise and heterogeneity, and support evolution in the formats over time.

\paragraph{Wrapper induction}
The goal of wrapper induction is to generate a set of extraction rules from an
annotated HTML document. 
Wrapper induction is an active area of research with techniques based on
supervised
learning~\cite{SatpalBSRS11,ZhangAYJS15,ZhuNWZM05,ZhuNWZM06,ZhuNZW08}, program
synthesis~\cite{raza2020web,raza2017automated,raza2018}, unsupervised
data mining~\cite{KordomatisHFKHB13,ArasuG03,MeccaCM01,ZhaiL06}, and programming
by demonstration~\cite{adelberg1998nodose}.
While some of these works deal with HTML documents as a series of
tokens~\cite{hsu1998generating, kushmerick1997wrapper}, others can leverage the
DOM tree structure of HTML~\cite{myllymaki2002robust, sahuguet1999building}.
The main distinguishing feature of our work is that \lrsyn is a generic
framework that can be instantiated across varied data formats as opposed to being
restricted to HTML documents.
Wrapper induction literature has also examined the possibility of repairing
extraction rules when new data of different formats
arrives~\cite{kushmerick1999regression, raposo2005automatic, lerman2003wrapper,
chidlovskii2006documentum}.
\lrsyn, due to its robustness, alleviates the need to update extraction
rules very frequently.
However, extraction rules will still occasionally break when format changes
significantly, i.e., when the landmarks change or there are format changes
within the ROI.
In these cases, incremental learning through repair is an interesting direction
to explore in the future.

The closest wrapper induction work to ours is Muslea et al~\cite{muslea1999hierarchical}, where the
authors use landmarks in a spirit similar to \lrsyn.
In~\cite{muslea1999hierarchical}, the wrapper induction rules go directly from
the landmark to the field value using a path of parent-child relations where
each step in the path specifies the type of DOM node.
In contrast, our technique has an intermediate step---we go from the landmark to
the ROI and from the ROI to the field value.
This affords us greater flexibility in the class of synthesis techniques that
can be used for extracting the field value from the ROI rather than just
following a fixed sequence of parent-child hops.
Another point of interest is that we use both parent hops and sibling hops in
going from the landmark to the region---this simplifies the extraction task by
allowing a larger class of landmark candidates.

\paragraph{Web testing and automation}
The web testing and automation domain and HTML data extraction share a number of
techniques related to robustly identifying a DOM element (field value in data
extraction and element under test for web automation)
In web testing, we want to robustly specify actions in the testing script---for
example, rather than specifying ``click the button at XPath
body[2]/table[4]/tr[1]/td[2]'' we want to specify ``locate the text label
\emph{Complete purchase} and click the button near it''.
The former is not robust to changes in the global page format while the latter
is.
Several tools, both research and commercial, provide web developers a high level
language or framework to write robust automation scripts
~\cite{LeshedHML08,LiNLDC10,LinWNCL09,imacros,Selenium,BeautifulSoup}.
Some of these tools even allow the developer to use annotation or labeling to
specify DOM elements from which scripts are automatically generated~\cite{LeshedHML08, imacros, LinWNCL09}.
Most web automation techniques deal with a single annotated document for
training instead of a full collection like in the data extraction
setting---hence, they do not deal with the heterogeneity problem.
The closest related work in this domain is Yandrapally et al~\cite{yandrapally2014robust} where
landmarks are inferred as an unambiguous ancestor of the target DOM node.
However, a node that is unambiguous for one document is not necessarily so
across multiple heterogeneous documents.
%
%
In fact, the $2$ level clustering and landmark inference is crucial to the \lrsyn
framework, while these steps are irrelevant in the web automation scenario.
Another difference between \lrsyn and \cite{yandrapally2014robust} is that we use a full-fledged DSL
and synthesizer for extracting the field value from the region which facilitates
more extractions on regions, as opposed to the ``near'' operator.
Another related work is~\cite{barman2016ringer}, where the authors use a node
similarity based addressing technique to identify nodes corresponding to
``labelled'' DOM nodes.
Here, the authors do not use any document structure at all and instead, locate
the target as the node that is most similar to the labelled node according a
weighted measure of the attributes common to both nodes.
This manner of node identification while robust to a certain
extent, fails when
the identification is over a set of heterogeneous documents from different sources.

\paragraph{Machine learning}
ML-based data extraction techniques have been explored in both the web and document image extraction domains, ranging from neural networks \cite{fb2020,lg2021,gs2020,cq2020,xl2020}, probabilistic models  \cite{zhang2015, dalvi2011} and markov logic networks \cite{satpal2011}. In general, ML approaches train opaque models that are neither readable nor editable by the user. One exception is the area of wrapper induction \cite{kushmerick2000} which learns sets of XPath expressions to handle noise \cite{dalvi2011,long2012, furche2016robust, LeottaSRT16}. Due to the global nature of XPaths, when applied to large heterogeneous datasets, such approaches can lead to a large number of expressions in order to cover irrelevant variations in document formats. In contrast, our technique is local and modular, and results in smaller programs in both size and number. Our technique also applies to general DSLs in different domains rather than just XPaths for web extraction. Ideas around exploiting compositionality and data invariance have also been explored in previous works: ~\cite{brin1998extracting, agichtein2000snowball}~use commonly reoccurring phrasal patterns for web extraction given a seed set; in the vision community, modular approaches such as convolutional neural networks have been used for document image extraction  \cite{fb2020,lg2021,xl2020,ss2019,zp2020}, and notably algorithms based on R-CNN~\cite{girshick2014rich}~  use selective search to focus attention on a small number of regions from the image (\emph{region proposals}). Our core ideas are similarly based around localised regions, but we detect them by identifying landmarks that present a common kind of invariance in formed documents. 

\paragraph{Hybrid approaches}
Recent approaches have combined program synthesis and ML techniques for data extraction. The closest related work in this area is \cite{ij2019}, where an ML model is used to get an initial labeling of potential attribute values, and the noisy labels produced by this model are used to create interpretable programs using synthesis techniques. While this approach shows improved robustness, it still generates global programs that can fail with irrelevant changes to the document format, and we show in this work how our compositional synthesis approach  performs better empirically in practice. There has been very limited work in the area of synthesis for document image extraction, but notable works in specialized areas include \cite{ss2019}, where concepts from inductive logic programming are combined with neural approaches, and \cite{zp2020}, which combines symbolic reasoning with CNNs, though interpretable programs are not generated. 

\leaveout{
\paragraph{Program synthesis for data extraction}
Data extraction has been an active area of investigation in the program synthesis community in recent years. Flashfill \cite{g2011}, which has been shipped in Microsoft Excel, generates small string processing programs in spreadsheets from examples. FlashExtract \cite{lg2014} synthesizes programs from examples for extraction from large text or web documents, using an algebra of pre-defined operators such as map, reduce and filter. These approaches work well when inputs are homogeneous, but for heterogeneous formats it is challenging to generate programs that cover all
inputs using only a few training examples. Several works have explored \emph{disjunctive} synthesis approaches to handle varying requirements for different subsets of inputs \cite{alur2015, alur2017, reynolds2015, saha2015, raza2018}, and \emph{Forgiving XPaths} ~\cite{omari2017synthesis}~ in particular focuses on synthesizing progressively relaxed XPaths to increase recall for web extraction. But such approaches do not focus on localised regions around the extraction and can therefore either  overgeneralize and take a hit on precision, or overfit to the global document formats. Hybrid synthesis ~\cite{raza2020web} proposes a combination of deductive \cite{pg2015} and predictive ~\cite{raza2017automated}~ synthesis techniques to create programs that follow aligned structures inferred on webpages, but this is primarily focused on tabular data extraction. Raychev et al. \cite{raychev2016} also learn programs from noisy data, but generate a single program from a
noisy dataset rather than using clustering to learn different programs applicable to different region formats. In contrast to the above techniques, we propose the idea of landmarks and regions as a compositional approach that addresses the high noise and heterogeneity in formats by inferring extraction within localized regions in the document, where such regions can be orthogonally detected using landmarks and clustering techniques that leverage data invariance across such documents.



\paragraph{Machine learning approaches for extraction}

\mrsays{do we want to claim ML approaches require significantly more training data than ours? not sure if we can make that point in terms of how many examples we require, and few shot is becoming popular in ML these days too }

Extracting attributes from heterogeneous data has been well-studied in the machine learning and data mining communities, where the aim is to train an ML model that can be used to predict extractions across heterogeneous document formats. Such models have been explored in both the web and document image extraction domains, and range across different techniques including neural networks \cite{12,fb2020,lg2021,gs2020,cq2020,xl2020}, probabilistic models  \cite{zhang2015, dalvi2011,long} and others such as markov logic networks \cite{satpal2011}. In general, ML approaches train complex models that lack interpretability and do not infer programs that are readable or editable by the user. One exception is the area of wrapper induction \cite{kushmerick2000} using ML techniques, such as learning sets of XPath expressions to handle noise \cite{dalvi2011,long2012}. But due to the global nature of XPaths, in practice these can again lead to huge  sets of  expressions in order cover irrelevant variations in document formats that can become difficult to understand, in contrast to the compositional programs we generate that concisely capture the relevant localized region variations only. Our technique also applies to general DSLs in different domains rather than just XPaths for web extraction. Interestingly, similar ideas of composisionality can be observed in document image extraction approaches that use CNNs \cite{fb2020,lg2021,xl2020,ss2019,zp2020}, which also take advantage of the hierarchical pattern in data.
Our core ideas are similar but formulated within the program synthesis setting, with the key distinction that it produces an interpretable and editable program artifact based on the compositional notions of landmarks and regions.







\paragraph{Approaches combining program synthesis and ML}

Interestingly, some recent approaches have explored combinations of program synthesis and machine learning techniques. The closest related work in this area is \cite{ij2019}, where an ML model is used to get an initial labeling of potential attribute values, and the possibly noisy labels produced by this model are used to create interpretable programs using synthesis techniques. While this approach shows improved robustness, it still generate global programs that can fail with irrelevant changes to the document format, and we show how our compositional synthesis approach  performs better empirically in practice. In fact, our compositionality contribution is orthogonal to the idea of using ML models, and we can explore augmentations of our system with ML models as in \cite{ij2019} with possible further improvements. There has been very limited work in the area of program synthesis for document image extraction, but some notable works in specialized areas include \cite{ss2019}, where concepts from inductive logic programming are combined with neural approaches, and \cite{zp2020}, which combines symbolic reasoning techniques with CNNs, though not with the purpose of generating interpretable programs. We bring mainstream synthesis techniques to the domain of data extraction from document images, and also show how this domain fits as an instantiation of our abstract landmark-based extraction approach that is also applicable in other extraction domains.

}

\vspace{-1ex}
\section{Conclusion}
\label{sec:conclusion}



Inspired by how humans search for data in formed documents, we designed a new approach to data extraction using the concepts of landmarks and regions. Our implementation of this approach, \lrsyn, is robust to format changes, and achieves close to perfect F1 scores with HTML documents, and significantly high F1 scores in image documents when compared to existing approaches. \lrsyn\ shines especially when test data has different formats than training data, which is a common pain-point in real-world applications.

While \lrsyn\ is robust to format changes that occur outside the region of interest, we believe that we can further improve its robustness for format changes inside the regions of interest by combining it with ML approaches as in \cite{ij2019}.
This is a very promising direction to explore due to the possibility of using smaller ML models trained directly on the regions of interest rather the the whole document.
Additionally, in our image datasets, we encountered documents where there are no specific landmark phrases present, though the regions of interest have similar visual structure. Using R-CNN models to come up with visual rather than textual landmarks could improve the performance of extraction in such cases.
\balance

\bibliography{references}


\end{document}